\documentclass[12pt,a4j]{article}
\usepackage[dvips]{graphicx}
\usepackage{amsmath}
\usepackage{amssymb}
\usepackage{enumerate}
\usepackage{amscd}
\usepackage{mathrsfs}

\begin{document}
\baselineskip=6mm

\centerline{\bf Integrability, conservation laws and solitons of a many-body dynamical system}\par
\centerline{\bf associated with the half-wave maps equation}\par 
\bigskip
\centerline{Yoshimasa Matsuno\footnote{{\it E-mail address}: matsuno@yamaguchi-u.ac.jp}}\par
\centerline{\it Division of Applied Mathematical Science, } \par
\centerline{\it Graduate School of Sciences and Technology for Innovation,}
\centerline{\it Yamaguchi University, Ube, Yamaguchi 755-8611, Japan} \par
\bigskip
\bigskip
\bigskip
\noindent {\bf Abstract} \par
\noindent  We consider the half-wave maps (HWM) equation which is a continuum limit of  the classical version of the Haldane-Shastry spin chain.
In particular, we explore a many-body dynamical system arising from the HWM equation under the pole ansatz. The system is shown to be completely integrable by
demonstrating  that it exhibits a Lax pair and  relevant conservation lows.  Subsequently, the analytical multisoliton solutions of
the HWM equation are constructed by means of the pole expansion method.  The properties of the one- and two-soliton solutions
are then investigated in detail as well as their pole dynamics.  Last, an asymptotic analysis of the $N$-soliton solution reveals that no phase shifts appear
after the collision of solitons. 
This intriguing feature is  worth noting since it is the first example observed in the head-on collision of rational solitons.
A number of  problems remain open for the HWM equation, some of which are discussed in concluding remarks.
\par

\newpage
\leftline{\bf  1. Introduction} \par
The half-wave maps (HWM) equation arises from a continuum limit of a classical version of the Haldane-Shastry spin chain [1-3].  The latter is also  known as the
classical spin Calogero-Moser (CM)  system whose complete integrability has been established [4-6].
The HWM equation describes the time evolution of a spin density  ${\bf m}(x, t) \in \mathbb{S}^2$,  where $t$ and $x$  are the temporal and spatial variables, respectively
and  $\mathbb{S}^2$ is the two-dimensional (2D) unit-sphere.
The evolution equation for ${\bf m}$  is given by [1-3]
$${\bf m}_t={\bf m}\times H{\bf m}_x, \eqno(1.1a)$$
with  the nonlocal operator $H$ defined by
$$H{\bf m}(x, t)={1\over \pi}P\int_{-\infty}^\infty{ {\bf m}(y, t)\over y-x}dy, \eqno(1.1b)$$
being the Hilbert transform. The subscripts $t$ and $x$ appended to ${\bf m}$  denote partial derivative
and the symbol "$\times$" is the vector product of 3D vectors in $\mathbb{R}^3$ or $\mathbb{C}^3$. 
Specifically, the latter is defined by 
${\bf a}\times {\bf b}=(a_2b_3-a_3b_2, a_3b_1-a_1b_3, a_1b_2-a_2b_1)$ for  3D vectors ${\bf a}=(a_j)_{1\leq j\leq 3}$ and ${\bf b}=(b_j)_{1\leq j\leq 3}$.
Throughout the paper, we restrict our consideration to the analysis on the real line.  
\par
Our main concern is the complete integrability of the HWM equation since it has been derived from  a continuum limit of an integrable
system.  A recent study reveals that the HWM equation admits a Lax representation, as well as an infinite number of conservation laws [7, 8].
As is well-known, the integrable soliton equations exhibit multisoliton solutions.  The one-soliton (or traveling solitary wave) solution has been
obtained [1, 2] together with numerical computation of the two- and three-soliton solutions [1]. 
Very recently, a procedure which is now known as the pole expansion method [9] was  applied to the HWM equation [10].  
 The  method works well in obtaining the
rational solutions of certain nonlinear evolution equations such as the Korteweg-de Vries (KdV)  and  Benjamin-Ono (BO) equations [11-14] for which
the  equations of motion for the poles are governed by the finite-dimensional dynamical systems.  The treatment of solutions expressed by hyperbolic functions becomes more sophisticated
since the number of poles becomes infinite [9].
Although some numerical computations were performed to obtain soliton solutions of the HWM equation,  the detail of the interaction process of solitons
was not clarified [10]. One reason for this
 is that the explicit forms of the multisoliton solutions are not available  yet even for the two-soliton case. \par
 First, we summarize the main result given in [10] for future uses (see $\bf Theorem\, 2.1$ in [10]):  We introdece the pole ansatz for solutions of Eq. (1.1)
 $${\bf m}(x, t)={\bf m}_0+{\rm i}\sum_{j=1}^N{{\bf s}_j(t)\over x-x_j(t)}-{\rm i}\sum_{j=1}^N{{\bf s}_j(t)^*\over x-x_j(t)^*}, \eqno(1.2)$$
 where ${\bf m}_0$  is an arbitrary constant  vector in $\mathbb{S}^2$ describing the boundary value ${\bf m}(\pm\infty, t)$ at spatial infinity, 
 the complex functions  $x_j(t)$ represent poles in the upper-half complex plane $\mathbb{C}_+$,  ${\bf s}_j$ are spin variables which take values in $\mathbb{C}^3$,
  the asterisk denotes complex conjugate and $N$ is an arbitrary positive integer characterizing the total number of poles (or solitons).
   It follows by taking the scalar product of Eq. (1.1) with ${\bf m}$ that ${\bf m}\cdot{\bf m}_t=0$, and hence ${\bf m}^2$ is a constant independent of $t$, 
    which  is set to 1 hereafter.  This implies that one can put ${\bf m}_0^2=1$ as well.
 Then, the spin field  ${\bf m}$ from (1.2) solves  Eq. (1.1) if ${\bf s}_j$ and $x_j$ obey the system of nonlinear ordinary differential equations
 $$\dot{{\bf s}}_j(t)=-2\sum_{k\not=j}^N{{\bf s}_j(t)\times{\bf s}_k(t)\over (x_j(t)-x_k(t))^2}, \quad (j=1, 2, ..., N),\eqno (1.3)$$
 $$\ddot{x}_j(t)=4\sum_{k\not=j}^N{{\bf s}_j(t)\cdot{\bf s}_k(t)\over (x_j(t)-x_k(t))^3}, \quad (j=1, 2, ..., N),\eqno (1.4)$$
 with the initial conditions ${\bf s}_j(0)={\bf s}_{j,0}, x_j(0)=x_{j,0}$ and
 $$\dot{x}_j(0)={{\bf s}_{j,0}\times {\bf s}^*_{j,0}\over {\bf s}_{j,0}\cdot{\bf s}^*_{j,0}}\cdot \left({\rm i}{\bf m}_0-\sum_{k\not=j}^N{{\bf s}_{k,0}\over x_{j,0}-x_{k,0}}
 +\sum_{k=1}^N{{\bf s}^*_{k,0}\over x_{j,0}-x^*_{k,0}}\right), \quad (j=1, 2, ..., N). \eqno(1.5)$$
 Furthermore, the $2N$ constraints are imposed on ${\bf s}_{j,0}$ and  $x_{j,0}$ such that
 $${\bf s}_{j,0}^2=0, \quad {\bf s}_{j,0}\cdot \left({\rm i}{\bf m}_0-\sum_{k\not=j}^N{{\bf s}_{k,0}\over x_{j,0}-x_{k,0}}
 +\sum_{k=1}^N{{\bf s}^*_{k,0}\over x_{j,0}-x^*_{k,0}}\right)=0, \quad (j=1, 2, ..., N). \eqno(1.6)$$
 In the above expressions, the notation $\sum_{k\not=j}^N$  is  short for the sum $\sum_{\substack{k=1\\ (k\not=j)}}^N$ and the dot appended to ${\bf s}_j$ denotes the time derivative. 
 The scalar product of 3D vectors ${\bf a}=(a_j)_{1\leq j\leq 3}$ and ${\bf b}=(b_j)_{1\leq j\leq 3}$ in $\mathbb{C}^3$ is defined by ${\bf a}\cdot{\bf b}=\sum_{j=1}^3a_jb_j$.
 Recall that the spin variables ${\bf s}_j$ and poles  $x_j$ in Eqs. (1.3) and (1.4) evolve according to the dynamics of an exactly solvable spin CM system which has been
 explored extensively in [4-6].  However, due to the constraints (1.6), the analysis of solutions becomes more complicated than that of the original spin CM system.
 It is important to remark that the constraints (1.6) hold for arbitrary $t>0$ if they are satisfied at $t=0$ [10].  In particular, taking the scalar product of ${\bf s}_j$  with
 Eq. (1.3) leads to the relation ${\bf s}_j\cdot \dot{\bf s}_j=(1/2)({\bf s}_j^2)_t=0$. Thus,   ${\bf s}_j^2(t)=0, \,(j=1, 2, ..., N)$ for arbitrary $t$.
 Note that since ${\bf s}_j\in \mathbb{C}^3$, the relation ${\bf s}_j^2={\bf s}_j\cdot{\bf s}_j=0$ does not imply ${\bf s}_j={\bf 0}$.
 Unlike the spin CM case, the permissible solutions are restricted by these conditions.
 \par
 The purpose of the present paper is twofold. The first one is to explore a many-body dynamical system associated with the HWM equation which has been described above.
 In particular, we show that it admits a  Lax pair as well as a number of conservation laws, showing that the system is completely integrable.
 Although it has been pointed out in [10] that the dynamical system governed by the equations of motion (1.3) and (1.4) is identical with the rational spin CM system,
 the explicit form of the Lax pair has not been discovered yet.
 The second one is to provide the general $N$-soliton formula of the HWM equation under the pole ansatz (1.2).   \par
 The present paper is organized as follows.
   In Section 2, we present a Lax pair for  equations (1.3) and (1.4). 
   In Section 3, we  derive the conservation laws of our system by using the Lax pair and exemplify some of them.  We also clarify the Hamiltonian structure of the system, and
   subsequently, the integration of the system   is performed by means of the standard procedure [15].
   In Section 4, we  develop an exact method of solution for constructing the multisoliton solutions of the
 HWM equation.  We use a procedure employed in [16] to obtain the rational multisoliton solutions of the nonlocal nonlinear Schr\"odinger (NLS) equation.
 In Section 5, we give  the explicit forms of  the one- and two-soliton solutions. 
Specifically,  the interaction process of two solitons is investigated in detail based on  their pole dynamics.  
The feature of the $N$-soliton solution is discussed briefly by focusing on its asymptotic behavior.
Section 6 is devoted to concluding remarks in which some open problems associated with the HWM equation are addressed.  
In Appendices A-D, the four propositions posed in Section 5 are proved. \par
\bigskip
\noindent {\bf 2. Lax pair and integration of the system of equations}\par
\bigskip
\noindent {\it 2.1. Lax pair}\par
\bigskip
Here, we establish the following theorem.\par
\bigskip
\noindent {\bf Theorem 1.} {\it The system of equations  (1.3) and (1.4) for ${\bf s}_j$ and $x_j$  admits a Lax pair
$$\dot{L}=[B, L]\equiv BL-LB, \eqno(2.1)$$
where $L$ and $B$ are $N\times N$ matrices whose elements are given respectively  by
$$L=(l_{jk})_{1\leq j,k\leq N}, \quad l_{jk}=\delta_{jk}\dot{x}_j+(1-\delta_{jk}){\epsilon_{jk}\sqrt{2\,{\bf s}_j\cdot{\bf s}_k}\over x_j-x_k}, \eqno(2.2)$$
$$B=(b_{jk})_{1\leq j,k\leq N}, \quad b_{jk}=(1-\delta_{jk}){\epsilon_{jk}\sqrt{
2\,{\bf s}_j\cdot{\bf s}_k}\over (x_j-x_k)^2}. \eqno(2.3)$$
Here, $\delta_{jk}$ is Kronecker's delta and $\epsilon_{jk}$ is an anti-symmetric symbol defined by
$$\epsilon_{jk}=-\epsilon_{kj},\quad \epsilon_{jk}^2=1-\delta_{jk}, \quad (j, k=1, 2, ..., N). \eqno(2.4)$$
}
\noindent {\bf Proof.} First, we provide a key relation which will be used frequently in our analysis:
$$({\bf s}_j\times{\bf s}_k)\cdot{\bf s}_l=\epsilon_{jk}\epsilon_{kl}\epsilon_{lj}\sqrt{2
}\sqrt{{\bf s}_j\cdot{\bf s}_k}\sqrt{{\bf s}_k\cdot{\bf s}_l}\sqrt{{\bf s}_l\cdot{\bf s}_j}. \eqno(2.5)$$
To verify (2.5), we note the identity
$$\{({\bf s}_j\times{\bf s}_k)\cdot{\bf s}_l\}^2=2({\bf s}_j\cdot{\bf s}_k)({\bf s}_k\cdot{\bf s}_l)({\bf s}_l\cdot{\bf s}_j), \eqno(2.6a)$$
which follows from the formula
$$\{({\bf s}_j\times{\bf s}_k)\cdot{\bf s}_l\}^2=\begin{vmatrix}{\bf s}_l^2&{\bf s}_j\cdot{\bf s}_l&{\bf s}_k\cdot{\bf s}_l \\
                                                                {\bf s}_j\cdot{\bf s}_l&{\bf s}_j^2&{\bf s}_j\cdot{\bf s}_k \\
                                                                {\bf s}_k\cdot{\bf s}_l&{\bf s}_j\cdot{\bf s}_k&{\bf s}_k^2\end{vmatrix}, \eqno(2.6b)$$
and the constraints ${\bf s}_j^2={\bf s}_k^2={\bf s}_l^2=0$.
 The square root of $(2.6a)$ yields, after taking into account the properties of the scalar and vector products, (2.5). \par
 To establish (2.1), we compute the $(j, k)$ elements of both sides to obtain
 $$(\dot{L})_{jk}=\delta_{jk}\ddot{x}+(1-\delta_{jk})\left\{-{\epsilon_{jk}\sqrt{2\,{\bf s}_j\cdot{\bf s}_k}\over (x_j-x_k)^2}(\dot{x}_j-\dot{x}_k)
 +{\epsilon_{jk}\over \sqrt{2}(x_j-x_k)}{\dot{\bf s}_j\cdot{\bf s}_k+{\bf s}_j\cdot \dot{\bf s}_k\over \sqrt{{\bf s}_j\cdot{\bf s}_k}}\right\}, \eqno(2.7)$$
$$([B, L])_{jk}=-(1-\delta_{jk}){\epsilon_{jk}\sqrt{2\,{\bf s}_j\cdot{\bf s}_k}\over (x_j-x_k)^2}(\dot{x}_j-\dot{x}_k)
  +2\sum_{l\not=j, k}^N{\epsilon_{jl}\epsilon_{lk}\sqrt{{\bf s}_j\cdot{\bf s}_l}\sqrt{{\bf s}_l\cdot{\bf s}_k}\over(x_j-x_l)^2(x_k-x_l)^2}(2x_l-x_j-x_k). \eqno(2.8)$$
  For $j=k$, noting the relation $\epsilon_{jl}\epsilon_{lj}=-1, \,(j\not=l)$, Eq. (2.1) with (2.7) and (2.8)  yields (1.4) whereas for $j\not=k$, it reduces to
  $${\epsilon_{jk}\over \sqrt{2}(x_j-x_k)}{\dot{\bf s}_j\cdot{\bf s}_k+{\bf s}_j\cdot \dot{\bf s}_k\over \sqrt{{\bf s}_j\cdot{\bf s}_k}}
  =2\sum_{l\not=j, k}^N{\epsilon_{jl}\epsilon_{lk}\sqrt{{\bf s}_j\cdot{\bf s}_l}\sqrt{{\bf s}_l\cdot{\bf s}_k}\over(x_j-x_l)^2(x_k-x_l)^2}(2x_l-x_j-x_k). \eqno(2.9)$$
Multiplying $\epsilon_{jk}\sqrt{2}(x_j-x_k)\sqrt{{\bf s}_j\cdot{\bf s}_k}$ on both sides of (2.9) and using the relations (2.5) and $\epsilon_{jk}^2=1, \,(j\not=k)$,
Eq. (2.9) can be recast in the form
$$\left\{\dot{\bf s}_j+2\sum_{l\not=j}^N{{\bf s}_j\times{\bf s}_l\over (x_j-x_l)^2}\right\}\cdot{\bf s}_k
+\left\{\dot{\bf s}_k+2\sum_{l\not=k}^N{{\bf s}_k\times{\bf s}_l\over (x_k-x_l)^2}\right\}\cdot{\bf s}_j=0. \eqno(2.10)$$
Since ${\bf s}_j$ and ${\bf s}_k$ satisfy Eq. (1.3), the left-hand side of (2.10) becomes zero, which proves the Lax Eq. (2.1).
 \quad $\Box$ \par
 We note that the system of equations (1.3) for ${\bf s}_j$ has a Lax representation
 $$\dot{S}=[B, S], \eqno(2.11)$$
 where $S$ is an $N\times N$ matrix with elements
 $$ S=(s_{jk})_{1\leq j,k\leq N}, \quad s_{jk}=\epsilon_{jk}\sqrt{2\,{\bf s}_j\cdot{\bf s}_k}. \eqno(2.12)$$
 The proof of (2.11) can be carried out in the same way as that of  Eq. (2.1).  Actually, Eq. (2.11) reduces to (2.10) and hence it holds identically by virtue of   Eq. (1.3).
 The Lax pair (2.1) is formally identical with that of the spin CM system given in [4, 5].  
However,  missing the relation (2.5),  its
explicit form has not been found yet and presented here for the first time. \par
\bigskip          
\noindent {\it 2.2. Integration of the system of equations}\par
\bigskip
To integrate the system of equations (1.3) and (1.4),  we provide the following proposition:\par
\bigskip
\noindent{\bf Proposition 1.}  {\it Let $X$ be an $N\times N$ matrix with elements
$$X=(x_{jk})_{1\leq j,k\leq N}, \quad x_{jk}=\delta_{jk}x_j. \eqno(2.13)$$
Then, $X$  evolves according to the equation 
$$\dot{X}=L+[B, X]. \eqno(2.14)$$
}
\bigskip
\noindent {\bf Proof.}  The proof can  be done    by a direct computation using  (2.2) 
and (2.3). \quad $\Box$ \par
Let us now solve the system of equations (1.3) and (1.4).  First, we introduce the quantity $J=J(t)=U^{-1}(t)X(t)U(t)$,  where $U$ satisfies the 
equation $\dot{U}=BU$  subjected to  the initial condition $U(0)=I \,(I: N\times N$  unit matrix).
It then follows from (2.1) and (2.14) that
$$\dot{J}=U^{-1}LU,  \eqno(2.15a)$$
$$\quad \ddot{J}=0. \eqno(2.15b)$$
Eq. $(2.15b)$ can be integrated twice with respect to $t$, giving
$$J(t)=\dot{J}|_{t=0}t+J(0)=L(0)t+X(0). \eqno(2.16)$$
The matrix $U(t)$  determines the time evolution of $L$ and $S$ in accordance with  the relations
$$L(t)=U(t)L(0)U(t)^{-1}, \eqno(2.17)$$
$$S(t)=U(t)S(0)U(t)^{-1}, \eqno(2.18)$$
thus providing a complete set of solutions to the system of equations (1.3) and (1.4). 
See also [4, 5] for more detailed discussion on the integration of the system under consideration.
\par
To proceed, we introduce the tau-function $f_N$ which plays a central role in constructing soliton solutions:
$$f_N=\prod_{j=1}^N(x-x_j)=|Ix-X|. \eqno(2.19)$$
Referring to the relation $X=UJU^{-1}$ with (2.16), we find that                                                                                                                                                                                                                                                       
\begin{align}              
   f_N&=|U(Ix-L(0)t-X(0))U^{-1}| \notag\\
    &=|Ix-L(0)t-X(0)|. \tag{2.20}
    \end{align}  
    The expressions of the poles $x_j$ can be obtained from (2.20) by solving the algebraic equation $f_N=0$  of the $N$th degree in $x$.
    However, in general, it is impossible to find their explicit analytical solutions.  In Section 4, we show that one needs only the
    fundamental symmetric polynomials of $x_j$  in constructing soliton solutions.  They  follow  immediately from (2.19) and (2.20)
    by means of a purely algebraic procedure.
    For latter use, we write the explicit forms of $f_1$ and $f_2$:
    $$f_1=x-\dot{x}_{1,0}t-x_{1,0}, \eqno(2.21a)$$
    $$f_2=x^2-\{(\dot{x}_{1,0}+\dot{x}_{2,0})t+x_{1,0}+x_{2,0}\}x+\left(\dot{x}_{1,0}\dot{x}_{2,0}-{2\,{\bf s}_{1,0}\cdot{\bf s}_{2,0}\over (x_{1,0}-x_{2,0})^2}\right)t^2$$
    $$+(\dot{x}_{1,0}x_{2,0}+x_{1,0}\dot{x}_{2,0})t+x_{1,0}{x}_{2,0}. \eqno(2.21b)$$
    \par
    The construction of the solutions for ${\bf s}_j$,  on the other hand,  is not addressed here. It will be considered in Section 4 where 
     we develop a new method of solution based on an elementary theory of linear algebra. \par
     \bigskip \noindent {\bf Remark 1.}\ In the case of the periodic solutions, the pole ansatz may be expressed in the form [10]
     $${\bf m}(x, t)={\bf m}_0+{\rm i}\sum_{j=1}^N{\bf s}_j(t)\kappa \cot\,\kappa\left(x-x_j(t)\right)
       -{\rm i}\sum_{j=1}^N{\bf s}^*_j(t)\kappa \cot\,\kappa\left(x-x_j^*(t)\right), \eqno(2.22)$$
     where $\kappa$ is a positive parameter. Then, the evolution equations of ${\bf s}_j$ and $x_j$ read [10]
     $$\dot{\bf s}_j(t)=-2\sum_{k\not=j}^N{\bf s}_j(t)\times{\bf s}_k(t)\,{\kappa^2\over \sin^2[\kappa\left(x_j(t)-x_k(t)\right)]}, \quad (j=1, 2, ..., N), \eqno(2.23)$$
     $$\ddot{x}_j(t)=4\sum_{k\not=j}^N{\bf s}_j(t)\cdot{\bf s}_k(t)\,{\kappa^3\,\cos[\kappa\left(x_j(t)-x_k(t)\right)]\over \sin^3[\kappa\left(x_j(t)-x_k(t)\right)]},
      \quad (j=1, 2, ..., N). \eqno(2.24)$$
      \par
       The Lax pair for the above system of equations takes the same  form as (2.1)  except that the elements of the matrices $L$ and $B$  are 
     given respectively by
     $$l_{jk}=\delta_{jk}\dot{x}_j+(1-\delta_{jk})\epsilon_{jk}\sqrt{2\,{\bf s}_j\cdot {\bf s}_k}\,{\kappa\over \sin\,\kappa(x_j-x_k)}, \quad (1\leq j, k\leq N), \eqno(2.25)$$
     $$b_{jk}=(1-\delta_{jk})\epsilon_{jk}\sqrt{2\,{\bf s}_j\cdot {\bf s}_k}\,{\kappa^2\cos\kappa(x_j-x_k)\over \sin^2\,\kappa(x_j-x_k)}, \quad (1\leq j, k\leq N). \eqno(2.26)$$
     As confirmed easily, in the limit of $\kappa\rightarrow 0$, the expressions (2.22)-(2.26) reduce to the corresponding ones for the soliton solutions.  The analysis of the periodic solutions is more
     involved than that  of the soliton solutions. Recall that the similar periodic problem has been exploited  in [16] for the nonlocal NLS equation.  
     This interesting issue will be addressed elsewhere. \par
     \bigskip
     \medskip
     \noindent {\bf 3. Conservation laws and Hamiltonian formulation}\par
     \bigskip
    The Lax pairs (2.1) and (2.11) for $L$ and $S$ allow us to obtain the  conservation laws  for the dynamical system
    described by the equations of motion (1.3) and (1.4).  Here, we derive several independent conservation laws. We also show that our
    system of equations can be written as a Hamiltonian system under appropriate Poisson brackets. \par
    \bigskip
       \noindent {\it 3.1. Conservation laws}\par
    \bigskip
      The direct consequence of (2.1) and (2.11)  is given by the following proposition:\par
    \bigskip
    \noindent{\bf Proposition 2.} {\it The quantities
    $$\mathscr{H}={1\over n}\,{\rm Tr}(L+\mu S)^n, \eqno(3.1)$$
   are the constants of motion,  where $n$ is an arbitrary positive integer and $\mu$ is an expansion parameter.} \par
    \bigskip
    \noindent{\bf Proof.}  Using (2.1) and (2.11) with the trace identity ${\rm Tr}(AB)={\rm Tr}(BA)$
    \begin{align}
  \dot{\mathscr{H}} &={\rm Tr}\left\{(\dot{L}+\mu\dot{S})(L+\mu S)^{n-1}\right\} \notag \\
                       &={\rm Tr}\left\{[B, L+\mu S](L+\mu S)^{n-1}\right\} \notag \\
                       &={\rm Tr}\left\{B(L+\mu S)^n-(L+\mu S)B(L+\mu S)^{n-1}\right\} \notag \\
                       &={\rm Tr}\left\{B(L+\mu S)^n-B(L+\mu S)^n\right\}\notag \\
                       &=0. \tag{3.2}
                       \end{align}
    $\Box$ \par
    Expanding (3.1) in powers of $\mu$,  one can see that the coefficients  at different powers of $\mu$ are also conserved. 
   In particular,  the quantities
    $$\mathscr{H}_{m, n}={1\over m+n}{\rm Tr}(L^mS^n), \eqno(3.3)$$
    are the constants of motion.  Below, we present some explicit examples: \par
    $$\mathscr{H}_{2, 0}={1\over 2}\sum_{j=1}^N\dot{x}_j^2+\sum_{j\not=k}^N{{\bf s}_j\cdot{\bf s}_k\over (x_j-x_k)^2}, \eqno (3.4a)$$
    $$\mathscr{H}_{3, 0}={1\over 3}\sum_{j=1}^N\dot{x}_j^3+2\sum_{j\not=k}^N{\dot{x}_j{\bf s}_j\cdot{\bf s}_k\over (x_j-x_k)^2}
    +{2\over 3}\sum_{j\not=k\not=l}^N{{\bf s}_j\cdot({\bf s}_k\times {\bf s}_l)\over ((x_j-x_k)(x_k-x_l)(x_l-x_j)}, \eqno(3.4b)$$
    $$\mathscr{H}_{1, 1}=-\sum_{j\not=k}^N{{\bf s}_j\cdot{\bf s}_k\over x_j-x_k}, \eqno(3.4c)$$
    $$\mathscr{H}_{1, 2}=-{1\over 3}\,\sum_{j\not=k}^N\dot{x}_j{\bf s}_j\cdot{\bf s}_k+{2\sqrt{2}\over 3}\,\sum_{j\not=k\not=l}^N{{\bf s}_j\cdot({\bf s}_k\times {\bf s}_l)\over x_j-x_k}. \eqno(3.4d)$$
    \par
    \medskip
    As in the case of the CM system, we have additional constants of motion [17].  To show this, we define the quantity $\mathscr{I}_n={\rm Tr}(XL^{n-1})$. Then, we establish  \par
    \bigskip
     \noindent{\bf Proposition 3.}\par
     $$\dot{\mathscr{I}}_n={\rm Tr}\,L^n. \eqno(3.5)$$
     \bigskip
    \noindent {\bf Proof.} Using (2.1) and (2.14), one can perform a sequence of computations to obtain
    \begin{align}
    \dot{\mathscr{I}}_n&={\rm Tr}\left(\dot{X}L^{n-1}+X\sum_{j=2}^nL^{n-j}\dot{L}L^{j-2}\right) \notag \\
                     &={\rm Tr}\left((L+[B, X])L^{n-1}+X\sum_{j=2}^nL^{n-j}[B, L]L^{j-2}\right) \notag \\
                     &={\rm Tr}\left(L^n+BXL^{n-1}-XBL^{n-1}+X(BL^{n-1}-L^{n-1}B)\right) \notag \\
                     &={\rm Tr}\, L^n. \tag{3.6} 
                     \end{align}
                     $\Box$  \par
                     \medskip
       Integration of (3.5) gives
   \begin{align}
   \mathscr{I}_n&=t\,{\rm Tr}\,L^n(0)+{\rm Tr}\,L^n(0) \notag \\
                &=nt \mathscr{H}_{n, 0}+{\rm Tr}\,L^n, \tag{3.7}
     \end{align}
     where ${\rm Tr}\,L^n\equiv{\rm Tr}\,L^n(t)={\rm Tr}\,L^n(0)$.
     It follows from (3.7) that the quantities
     \begin{align}
     \mathscr{I}_{m, n}&\equiv m \mathscr{I}_n\mathscr{H}_{m,0}-n \mathscr{I}_m\mathscr{H}_{n,0} \notag\\
     &={\rm Tr}(XL^{n-1}){\rm Tr}\,L^m-{\rm Tr}(XL^{m-1}){\rm Tr}\,L^n, \tag{3.8}
     \end{align}
     are the constants of motion. 
     In particular, for $m=1$, (3.8) reduces to
     $$\mathscr{I}_{1, n}={\rm Tr}(XL^{n-1}){\rm Tr}\,L-{\rm Tr}\, X\,{\rm Tr}\, L^n. \eqno(3.9)$$
     \par
     The information of the conservation laws about the dynamical system under consideration does not provide directly that of the conservation laws of the HWM equation itself.
     However, a few conservation laws are available for the HWM equation which are associated with the global symmetries of the equation [1]. 
     For instance, the integral $\mathscr{J}=(1/2)\int_{-\infty}^\infty{\bf m}\cdot H{\bf m}_xdx$ corresponding to energy (or Hamiltonian)  is preserved in $t$.
     Substituting  ${\bf m}$ from (1.2) into this expression, we obtain
     $$\mathscr{J}=2\pi\sum_{j,k=1}^N{{\bf s}_j\cdot{\bf s}_k^*\over (x_j-x_k^*)^2}. \eqno(3.10)$$
     As will be expected from the existence of a Lax pair [7],  an infinite number of conservation laws may exist for the HWM equation which would establish 
     under appropriate conditions the complete 
     integrability of the equation. This interesting issue deserves a future investigation. \par
     \bigskip
    \noindent {\it 3.2. Hamiltonian formulation}\par
    \bigskip
     We start from the Hamiltonian from $(3.4a)$
     $$\mathscr{H}_{2, 0}={1\over 2}\sum_{j=1}^Np_j^2+{1\over 2}\sum_{j\not=k}^N{s_{jk}^2\over (x_j-x_k)^2}, \eqno(3.11)$$
     where $p_j=\dot{x}_j$ are the momentum variables and  $s_{jk}$  are given by (2.12). 
          The system described by the Hamiltonian (3.11) has been introduced in [4, 5].
     The number of the independent variables is found to be $2N+N(N-1)/2=N(N+3)/2$  by taking into account  the antisymmetric property of  the variable $s_{jk}$.
     In accordance with [5], we define the Poisson brackets
     $$\{f, g\}_p=\sum_{j=1}^N\left({\partial f\over \partial x_j}{\partial g\over \partial p_j}-{\partial g\over \partial x_j}{\partial f\over \partial p_j}\right), \eqno(3.12)$$
     $$\{s_{ij}, s_{kl}\}_s=\delta_{il}s_{kj}+\delta_{ik}s_{jl}+\delta_{jl}s_{ik}+\delta_{jk}s_{li}. \eqno(3.13)$$
     Then, the equations of motion for $x_j$ and $p_j$  can be written as
     $$\dot{x}_j=\{x_j, \mathscr{H}_{2, 0}\}_p=p_j, \eqno(3.14)$$
     $$\dot{p}_j=\{p_j, \mathscr{H}_{2, 0}\}_p=\sum_{k\not=j}^N{2s_{jk}^2\over (x_j-x_k)^3}, \eqno(3.15)$$
     wheres those of $s_{jk}$  are given by
     $$\dot{s}_{jk}=\{s_{jk}, \mathscr{H}_{2, 0}\}_s=-\sum_{l\not=j,k}^N\left({s_{jl}s_{lk}\over (x_j-x_l)^2}-{s_{jl}s_{lk}\over (x_k-x_l)^2}\right). \eqno(3.16)$$
     The expressions (3.15) with (3.14) coincide with (1.4), and (3.16)  are found to reduce to  (2.10) after a few manipulations. This result exhibits  the Hamiltonian
    structure of the system.  Although the independence of the various conserved quantities derived in Section 3.1 under the above Poisson brackets 
     is an interesting issue, we do not discuss it here and instead refer  to [4]. \par
     \bigskip
     \noindent{\bf 4. Construction of the $N$-soliton solution}\par
     \bigskip
     \noindent {\it 4.1. N-soliton solution}\par
     \bigskip
     Here, we provide an explicit formula for the rational $N$-soliton solution of the HWM equation.  
     It  has a form of the pole representation given by (1.2).
         To this end, we make a few preparations.  First, we write the tau-function $f_N$ from (2.19)  in the form
     $$f_N=\sum_{j=0}^N(-1)^j\sigma_jx^{N-j}, \eqno(4.1a)$$
     where $\sigma_j$ are fundamental symmetric polynomials of $x_j$ given by
     $$\sigma_0=1, \quad \sigma_1=\sum_{j=1}^Nx_j, \quad \sigma_2=\sum_{j<k}^Nx_jx_k, ...,\quad  \sigma_N=\prod_{j=1}^Nx_j. \eqno(4.1b)$$
     We define the polynomials $\sigma_{n,j}$ by the relation  
     $$(x-x_1)...(x-x_{j-1})(x-x_{j+1})...(x-x_N)=\sum_{n=0}^{N-1}(-1)^n\sigma_{n,j}x^{N-n-1}, \quad (j=1, 2, ..., N). \eqno(4.2)$$
     Multiplying $x-x_j$  by (4.2) and comparing the coefficients of $x^{N-n+l}$ on both sides, we obtain the recursion relation
     $\sigma_{n-l,j}+x_j\sigma_{n-l-1,j}=\sigma_{n-l}$. If we multiply $(-1)^lx_j^l$  by this and add the resultant expression from $l=0$ to $l=n-1$,  we arrive at the relation
     $$\sigma_{n,j}=\sum_{l=0}^n(-1)^lx_j^l\sigma_{n-l}. \eqno(4.3)$$
     Furthermore, we introduce the vector quantities
     $${\bf J}_n=\sum_{j=1}^N{\bf s}_jx_j^n, \quad (n=0, 1, 2, ...). \eqno(4.4)$$
     Then, we establish our main result:\par
     \bigskip
     \noindent {\bf Theorem 2.} {\it The solution  of the HWM equation under the pole ansatz (1.2) is  represented explicitly in terms of $\sigma_j$ and ${\bf J}_l$ as
     $${\bf m}={\bf m}_0
     +{\rm i}\left({\sum_{n=0}^{N-1}\sum_{l=0}^n(-1)^{n+l}{\bf J}_{\it l}\sigma_{n-l}x^{N-n-1}\over \sum_{j=0}^N(-1)^j\sigma_jx^{N-j}} -c.c.  \right), \eqno(4.5) $$   
     where the notation $c.c.$ stands for  the complex conjugate expression of the preceding expression.}\par
     \bigskip
     \noindent {\bf Proof.}  We modify ${\bf m}$ from (1.2) by using (4.2) to  obtain
     $${\bf m}={\bf m}_0+{\rm i}\left({1\over f_N}\sum_{j=1}^N\sum_{n=0}^{N-1}(-1)^n\sigma_{n,j}{\bf s}_jx^{N-n-1}- c.c\right).$$
     The $N$-soliton formula (4.5) follows simply by inserting (4.1), (4.3) and (4.4) into the above expression.
      \quad $\Box$ \par
     \bigskip
     \noindent {\bf Remark 2.} 
     Before proceeding, we make a few comments. First, the expressions (2.19) and (2.20) imply that $\sigma_n$ is an $n$th-order polynomial of $t$
     whose coefficients depend only on the initial conditions. As evidenced  by (4.5), the solutions $x_j$ themselves are not necessary,  but one needs the 
     polynomials $\sigma_j$ to obtain  ${\bf m}$. 
     This observation is quite important since the explicit analytical expression of  $x_j$ are not available in general. Second, the solutions ${\bf s}_j$ are expressed 
     in terms of $x_j\, (j=1, 2, ..., N)$ and ${\bf J}_n\, (n=1, 2, ..., N-1)$ 
     by solving the system of linear algebraic equations (4.4) for ${\bf s}_j$.
         The explicit formula for ${\bf s}_j$ will be presented in Section 4. See (4.30). We recall that an analogous formula has been given for 
     the rational $N$-soliton solution of the nonlocal NLS equation [16].          
     \par
     The following theorem is crucial in the subsequent analysis:\par
     \bigskip
      \noindent {\bf Theorem 3.}  {\it The  quantity ${\bf J}_n$ defined by (4.4) is an $n$th-order polynomial of $t$. More precisely, it can be expressed  in the form
      $${\bf J}_n=\sum_{k=0}^n{{\bf c}_{n,k}\over k!}\,t^k, \eqno(4.6a)$$
      with 
      $${\bf c}_{n,k}={d^k{\bf J}_n\over dt^k}\biggm|_{t=0}=\sum_{j=1}^N{d^k({\bf s}_jx_j^n)\over dt^k}\biggm|_{t=0}. \eqno(4.6b)$$}
      \par
      Note that the coefficients ${\bf c}_{n,k}$ are determined by the initial conditions for $x_j, \dot{x}_j$,  ${\bf s}_j$ and  $\dot{\bf s}_j\, (j=1, 2, ..., N)$.
      The expression (4.6) implies that $d^{n+1}{\bf J}_n/dt^{n+1}=0\, (n= 0, 1, ..., N-1)$ .
      A direct verification of these relations using (1.3) and (1.4) is possible for the first few $n$'s. However, the amount of computations becomes formidable as $n$ increases.
      Therefore, we employ an alternative approach. \par
      Theorem 3 is now  proved in several steps, which we shall now demonstrate by establishing  a sequence of propositions. \par
      \bigskip
      \noindent {\bf Proposition 4.}  {\it  Let
     $Y=(I-\epsilon X)^{-1}.$
      Then,
      $$\dot{Y}=BY-YB+\epsilon YLY, \eqno(4.7)$$
         where $\epsilon$ is an expansion parameter.} \par
      \bigskip
       \noindent {\bf Proposition 5.} {\it Let 
      $K=YL. $
       Then,
       $$\dot{K}=BK-KB+\epsilon K^2. \eqno(4.8)$$
       }
       \bigskip
       \noindent {\bf Proposition 6.} {Let 
       $\mathscr{P}_n={\rm Tr}(S^2K^nY). $
        Then,
       $$\dot{\mathscr{P}}_n=\epsilon (n+1) \mathscr{P}_{n+1}. \eqno(4.9)$$
       }
       \bigskip
       \noindent {\bf Proposition 7.} {Let 
       $\mathscr{Q}_n={\rm Tr}(S^2X^n). $
        Then,
      $$ \sum_{l=n}^\infty\epsilon^l\,{d^n\mathscr{Q}_l\over dt^n}=\epsilon^n n!\mathscr{P}_n. \eqno(4.10)$$
      \medskip
      Proposition 7 comes from Propositions 4-6  which are proved in Appendices A-D. \par
      Now, we are ready for the proof of Theorem 3. \par
      \bigskip
      \noindent{\bf Proof of Theorem 3.} 
      \ We rewrite  $\mathscr{Q}_n$ defined in (4.10) by using  (2.12) and (2.13), giving
            $$\mathscr{Q}_n=-2\left(\sum_{j=1}^N{\bf s}_jx_j^n\right)\cdot \left(\sum_{j=1}^N{\bf s}_j\right)=-2{\bf J}_n\cdot{\bf J}_0.  \eqno(4.11)$$
      We expand  $\mathscr{P}_n$ in powers of $\epsilon$ and compare
      the coefficient of $\epsilon^n$ on both sides of $(4.10)$.  This leads to  the relation
      $$ {d^n\mathscr{Q}_n\over dt^n} =n!{\rm Tr} (S^2L^n). \eqno(4.12)$$
      Differentiating (4.12) with respect to $t$ and  taking into account the constant of motion  (3.3), we obtain
      \begin{align}
      {d^{n+1}\mathscr{Q}_n\over dt^{n+1}}&=n!(n+2){d \mathscr{H}_{n, 2}\over dt} \notag \\
                                              &=0. \tag{4.13}
      \end{align}
      If we substitute (4.13) into (4.11), we find
      $${d^{n+1}{\bf J}_n\over dt^{n+1}}\cdot{\bf J}_0=0. \eqno(4.14)$$
      Since ${\bf J}_0=\sum_{j=1}^N{\bf s}_j$ is an arbitrary constant vector specified by the initial conditions,  one concludes that
      $d^{n+1}{\bf J}_n/dt^{n+1}=0$, and hence ${\bf J}_n$ is an $n$th-order  polynomial of $t$. This completes the proof of Theorem 3. \quad $\Box$ \par
      \bigskip
      \newpage
     \noindent {\it 4.2. Solution of the constraints}\par
     \bigskip
      For the complete description of  the $N$-soliton solution, we must specify the initial conditions for ${\bf s}_j$, $x_j$ and $\dot{x}_j$.
      This problem becomes complicated  because of the constraints (1.6) imposed on ${\bf s}_{j,0}      $.  
      We seek the solutions of (1.6) of the form [10]
      $${\bf s}_{j,0}=s_j({\bf n}_{j,1}+{\rm i}{\bf n}_{j,2}), \quad (j=1, 2, ..., N), \eqno(4.15)$$
      where  ${\bf n}_{j,1}$ and  ${\bf n}_{j,2} \in \mathbb{S}^2$ are unit vectors satisfying the conditions
      ${\bf n}_{j,1}\cdot{\bf n}_{j,2}=0\, (j=1, 2, ..., N)$
      and $s_j\in \mathbb{C}$ are unknown parameters to be determined later.
      In addition, we define the vectors
      $${\bf n}_{j,+}={\bf n}_{j,1}+{\rm i}{\bf n}_{j,2},\quad {\bf n}_j={\bf n}_{j,1}\times{\bf n}_{j,2}, \quad (j=1, 2, ..., N), \eqno(4.16)$$
      as well as the parameters
      $$\kappa_{jk}={\bf n}_{j,+}\cdot {\bf n}_{k,+}, \quad \nu_{jk}={\bf n}_{j,+}\cdot {\bf n}_{k,+}^*,
      \quad   \mu_j={\bf n}_{j,+}\cdot{\bf m}_0,\quad (j, k=1, 2, ..., N). \eqno(4.17)$$
      Note that ${\bf n}_j$ is a unit vector orthogonal to the vectors ${\bf n}_{j,1}$ and ${\bf n}_{j,2}$.
      The first constraints in (1.6) are fulfilled due to the orthogonality relations   ${\bf n}_{j,1}\cdot{\bf n}_{j,2}=0\, (j=1, 2, ..., N)$. 
      The second constraints, on the other hand, can be written in terms of the parameters defined by  (4.17) as
      $${\rm i}\mu_j+{2s_j^*\over x_{j,0}-x_{j,0}^*}-\sum_{k\not=j}^N{\kappa_{jk}s_k\over x_{j,0}-x_{k,0}}+\sum_{k\not=j}^N{\nu_{jk}s_k^*\over x_{j,0}-x_{k,0}^*}=0,
      \quad (j=1, 2, ..., N). \eqno(4.18)$$
      The complex conjugate expression of (4.18) reads
      $$-{\rm i}\mu_j^*+{2s_j\over x_{j,0}^*-x_{j,0}}-\sum_{k\not=j}^N{\kappa_{jk}^*s_k^*\over x_{j,0}^*-x_{k,0}^*}+\sum_{k\not=j}^N{\nu_{jk}^*s_k\over x_{j,0}^*-x_{k,0}}=0, 
      \quad (j=1, 2, ..., N). \eqno(4.19)$$
      If we introduce the matrices $F$ and $G$ together with the  vectors ${\bf s}$ and ${\boldsymbol \mu}$  whose elements are given by
      $$F=(f_{jk})_{1\leq j,k\leq N},\quad f_{jk}={2\delta_{jk}\over x_{j,0}-x_{k,0}^*}+(1-\delta_{jk})\,{\nu_{jk}\over x_{j,0}-x_{k,0}^*}, \eqno(4.20)$$
      $$G=(g_{jk})_{1\leq j,k\leq N}, \quad g_{jk}=-(1-\delta_{jk})\,{\kappa_{jk}\over x_{j,0}-x_{k,0}}, \eqno(4.21)$$
      $${\bf s}=(s_1, s_2, ..., s_N)^T, \quad {\bf s}^*=(s_1^*, s_2^*, ..., s_N^*)^T, \eqno(4.22)$$
      $${\boldsymbol \mu}=(\mu_1, \mu_2, .., \mu_N)^T,\quad {\boldsymbol \mu}^*=(\mu_1^*, \mu_2^*, .., \mu_N^*)^T,\eqno(4.23)$$
      then, the system of linear algebraic equations (4.18) and (4.19) for $s_j$ and $s_j^*$ can be put into the compact form as
       $$ \quad G{\bf s}+F{\bf s}^*=-{\rm i}{\boldsymbol \mu}, \eqno(4.24a)$$
      $$F^*{\bf s}+G^*{\bf s}^*={\rm i}{\boldsymbol \mu}^*.
       \eqno(4.24b)$$
     
      If the conditions $|F|\not=0$ and $|F^*-G^*F^{-1}G|\not=0$ are satisfied, then  Eqs. (4.24) can be solved uniquely for ${\bf s}$ to give
      $${\bf s}={\rm i(}F^*-G^*F^{-1}G)^{-1}(G^*F^{-1}{\boldsymbol \mu}+{\boldsymbol \mu}^*). \eqno(4.25)$$
      \par
     In the case of the two-soliton solution which will be described in detail in Section 5, (4.25) is expressed in the form
      $$s_1={\Delta_1\over \Delta}, \quad s_2={\Delta_2\over \Delta}, \eqno(4.26a)$$
      where
      $$\Delta=1+{\nu_{12}^*\nu_{12}\over 4}\,{(x_{1,0}-x_{1,0}^*)(x_{2,0}-x_{2,0}^*)\over |x_{1,0}^*-x_{2,0}|^2}-{\kappa_{12}^*\kappa_{12}\over 4}\,{(x_{1,0}-x_{1,0}^*)(x_{2,0}-x_{2,0}^*)\over |x_{1,0}-x_{2,0}|^2}, \eqno(4.26b)$$
      $$\Delta_1=-{{\rm i}\over 4}(x_{1,0}-x_{1,0}^*)(x_{2,0}-x_{2,0}^*)\left({\nu_{12}^*\mu_2^*\over x_{1,0}^*-x_{2,0}}-{\kappa_{12}^*\mu_2\over x_{1,0}^*-x_{2,0}^*}\right)
      -{{\rm i}\over 2}(x_{1,0}-x_{1,0}^*)\mu_1^*, \eqno(4.26c)$$
       $$\Delta_2=-{{\rm i}\over 4}(x_{1,0}-x_{1,0}^*)(x_{2,0}-x_{2,0}^*)\left({\nu_{12}\mu_1^*\over x_{2,0}^*-x_{1,0}}-{\kappa_{12}^*\mu_1\over x_{2,0}^*-x_{1,0}^*}\right)
      -{{\rm i}\over 2}(x_{2,0}-x_{2,0}^*)\mu_2^*. \eqno(4.26d)$$
            Note that $\Delta\in \mathbb{R}$ and $\kappa_{21}=\kappa_{12},\nu_{21}^*=\nu_{12}$. \par
      The solution of the constraints makes it possible to solve the initial value problem of the system of equations (1.3) and (1.4), which we shall summarize.
     For a given ${\bf m}_0$,  prepare the initial pole positions $x_{j,0}=x_j(0)\in {\mathbb C}_+\, (j=1, 2, ..., N)$ and the directions ${\bf n}_j\in {\mathbb S}^2$ of the initial
     spins ${\bf s}_{j,0}\, (j=1, 2, ..., N)$ which are given by (4.16).  The initial conditions for the spin variables  are computed in accordance with (4.15) and (4.25).  
     The initial conditions (1.5) for $\dot{x}_{j,0}\in {\mathbb C}$
     then  reduce, after introducing (4.15), to the transparent forms 
      $$\dot{x}_{j,0}=-{\rm i}{\bf n}_j\cdot \left({\rm i}{\bf m}_0-\sum_{k\not=j}^N{{\bf s}_{k,0}\over x_{j,0}-x_{k,0}}
 +\sum_{k=1}^N{{\bf s}^*_{k,0}\over x_{j,0}-x^*_{k,0}}\right), \quad (j=1, 2, ..., N). \eqno(4.27)$$
 \par
      \bigskip
      \noindent {\it 4.3. Consistency of an overdetermined system}\par
      \bigskip
      As already shown by (4.5),  the spin variables themselves are not required for the purpose of constructing
      the $N$-soliton solution. However, since these  are used to evaluate the asymptotic behavior of the solutions, we derive their explicit expressions.
      First, we note that the system of equations for ${\bf s}_j$ is overdetermined so that one must verify its consistency.
      Specifically, we show that it has a  solution. To this end,
 let ${\bf J}_{N+m}=\sum_{j=0}^{N-1}c_j{\bf J}_j\ (m\geq 0)$ with $c_j$ being unknown constants to be determined.  Invoking the definition of ${\bf J}_j$, this
expression can be written as $\sum_{k=1}^N{\bf s}_kx_k^{N+m}=\sum_{k=1}^N{\bf s}_k\sum_{j=0}^{N-1}c_jx_k^j$.
Equating the coefficients of ${\bf s}_k$ on both sides, one obtains the  system of linear algebraic equations 
for $c_j$: $\sum_{j=0}^{N-1}c_jx_k^j=x_k^{N+m}\ (k=1, 2, ..., N)$. This system is solved simply
to give the solution
$$c_j=(-1)^{N-j-1} \begin{vmatrix}R_0 &R_{-1} & \ldots &R_{-N+1} \\
                                 R_1& R_0 &\ldots & R_{-N+2} \\
                                 \vdots &\vdots& \ddots &\vdots \\
                                 R_{j-1}& R_{j-2}& \ldots& R_{-N+j} \\
                                 R_{j+1}& R_{j}& \ldots& R_{-N+j+2} \\
                                 \vdots &\vdots& \ddots &\vdots \\
                                 R_{N-1}& R_{N-2}& \ldots& R_{0} \\
                                   R_{N+m}& R_{N+m-1}& \ldots& R_{m+1}
                                   \end{vmatrix}, \quad (j=0, 1, 2, ..., N-1), \eqno(4.28a)$$
where $R_n$ are defined by the relation
$$\prod_{j=1}^N(1-\epsilon x_j)^{-1}=1+\sum_{n=1}^\infty \epsilon^n R_n, \eqno(4.28b)$$
with $R_0=1$ and $R_n=0\ (n<0)$.  It follows from $(4.28b)$ that the quantities $R_n$ are  
 expressed in terms of the fundamental symmetric polynomials $\sigma_j\ (j=1, 2, ..., N)$ defined by $(4.1b)$.
In the case of $N=3$ and $ m=0, 1, 2$, for instance, the resulting expressions of $ {\bf J}_3$, ${\bf J}_4$  and ${\bf J}_5$ are given respectively by
$${\bf J}_3=\sigma_1{\bf J}_2-\sigma_2{\bf J}_1+\sigma_3{\bf J}_0, \eqno(4.29a)$$
$${\bf J}_4=(\sigma_1^2-\sigma_2){\bf J}_2-(\sigma_1\sigma_2-\sigma_3){\bf J}_1+\sigma_1\sigma_3{\bf J}_0. \eqno(4.29b)$$
$${\bf J}_5=(\sigma_1^3-2\sigma_1\sigma_2+\sigma_3){\bf J}_2-(\sigma_1^2\sigma_2-\sigma_2^2-\sigma_1\sigma_3+\sigma_4){\bf J}_1
+(\sigma_1^2\sigma_3-\sigma_2\sigma_3-\sigma_1\sigma_4+\sigma_5){\bf J}_0. \eqno(4.29c)$$
\par
The above discussion reveals that  $N$ equations are independent among (4.4) and the other ones are redundant.  In accordance with this observation,  we apply Cramer's rule to
the first $N$ equations   and obtain the solution
 $$ 
 {\bf s}_j=\begin{vmatrix} 1&\ldots&1&{\bf J}_0 &1&\ldots& 1 \\
            x_1&\ldots &x_{j-1}& {\bf J}_1 & x_{j+1} &\ldots & x_N\\
            \vdots & \ddots&\vdots&\vdots&\vdots&\ddots&\vdots\\
            x_1^{N-1}&\ldots&x_{j-1}^{N-1}& {\bf J}_{N-1}& x_{j+1}^{N-1}& \ldots& x_N^{N-1}
            \end{vmatrix}/V_N, \quad (j=1, 2, ..., N), \eqno(4.30a)$$
 where
 $$V_N=|(x_k^{j-1})_{1\leq j,k\leq N}|=\prod_{1\leq k<j\leq N}(x_j-x_k), \eqno(4.30b)$$
 is the Vandermonde determinant. 
 If we expand the determinant in $(4.30a)$ with respect to the $j$th column, then we can express ${\bf s}_j$ as a linear combination of  ${\bf J}_k\, (k=0, 1, ..., N-1)$.
\par
\bigskip
\noindent{\bf 5.  Soliton solutions}\par
     \bigskip
     In this section, we present the explicit examples of soliton solutions. In particular, we explore the properties of the one- and two-soliton solutions which are most fundamental
     constituents among soliton solutions. The $N$-soliton solution will be described briefly focusing on its asymptotic behavior at large time. \par
              \bigskip
     \noindent {\it 5.1. One-soliton solution}\par
     \bigskip
     The one-soliton solution is given by (4.5) with $N=1$. It reads
     $${\bf m}={\bf m}_0+{\rm i}\left({{\bf s}_1(t)\over x-x_1(t)}-{{\bf s}_1^*(t)\over x-x_1^*(t)}\right), \eqno(5.1a)$$
     with
     $$ x_1(t)=\dot{x}_{1,0}\,t+x_{1,0}, \quad {\bf s}_1(t)={\bf s}_{1,0},\eqno(5.1b)$$
     where $(5.1b)$ follows from $(2.21a)$ and (4.6) with $n=0$. In accordance with (4.27), the initial condition of $\dot{x}_1$  is
     $$\dot{x}_{1,0}=-{\rm i}{\bf n}_1\cdot \left({\rm i}{\bf m}_0+{{\bf s}_{1,0}^*\over x_{1,0}-x_{1,0}^*}\right), \eqno(5.2a)$$
     and the constraint for  ${\bf s}_{1,0}$  comes from (1.6) with $N=1$, giving
     $${\bf s}_{1,0}\cdot\left({\rm i}{\bf m}_0+{{\bf s}_{1,0}^*\over x_{1,0}-x_{1,0}^*}\right)=0. \eqno(5.2b)$$
     If we substitute ${\bf s}_{1,0}$ from (4.15) into $(5.2b)$, we can determine $s_1$.  Consequently,
           $${\bf s}_{1,0}={\rm Im}\,x_{1,0}\,({\bf n}_{1,+}^*\cdot{\bf m}_0){\bf n}_{1,+}. \eqno(5.3)$$
     Introducing (5.3) into $(5.2a)$ and taking into account the relation ${\bf n}_{1,+}^*\cdot{\bf n}_1=0$, 
     we obtain $\dot{x}_{1,0}={\bf n}_1\cdot{\bf m}_0\equiv v\ (|v|<1)$.  It turns out from  $(5.1b)$ that
     $x_1(t)=vt+x_{1,0}$. With these results, the one-soliton solution (5.1) is expressed in the form
            $${\bf m}={\bf m}_0+{\rm i}\,{\rm Im}\,x_{1,0}\left\{{{\bf n}_{1,+}^*\cdot{\bf m}_0\over x-vt-x_{1,0}}\,{\bf n}_{1,+}-
            {{\bf n}_{1,+}\cdot {\bf m}_0\over x-vt-x_{1,0}^*}\,{\bf n}_{1,+}^*\right\}.\eqno(5.4)$$
            \par
     Let ${\boldsymbol e}_1=(1, 0, 0), {\boldsymbol e}_2=(0, 1, 0), {\boldsymbol e}_3=(0, 0, 1)$ be the normal orthogonal bases in ${\mathbb R}^3$.
     We then put ${\bf n}_{1,1}={\boldsymbol e}_1,{\bf n}_{1,2}={\boldsymbol e}_2, {\bf n}_1={\boldsymbol e}_3$ and  
            $${\bf m}_0=\cos\,\theta\,{\boldsymbol e}_1+\sin\,\theta\,{\boldsymbol e}_3, \eqno(5.5)$$
             where $\theta$ is a real parameter. Note in this setting that ${\bf n}_{1,+}={\boldsymbol e}_1+{\rm i}{\boldsymbol e}_2$.
     Using the relations $v={\bf n}_1\cdot{\bf m}_0=\sin\,\theta, {\bf n}_{1,+}^*\cdot{\bf m}_0=\cos\,\theta$, the expression of ${\bf m}$ from (5.4) is rewritten in the form
     \begin{align}
     {\bf m}&={\bf m}_0+{\rm i}\,{\rm Im}\,x_{1,0}\,\cos\,\theta\left({{\bf n}_{1,+}\over x-vt-x_{1,0}}-{{\bf n}_{1,+}^*\over x-vt-x_{1,0}^*}\right) \notag \\
            &={\bf m}_0-{2\,{\rm Im}\,x_{1,0}\,\cos\,\theta \over  (x-vt-{\rm Re}\,x_{1,0})^2+({\rm Im}\, x_{1,0})^2}
            \left\{{\rm Im}\, x_{1,0}\,{\boldsymbol e}_1+(x-vt-{\rm Re}\, x_{1,0})\,{\boldsymbol e}_2\right\}. \tag{5.6}
            \end{align}                  
            \par
     Introducing the function $B(z)$
     \begin{align}
     B(z)&={z-x_{1,0}\over z-x_{1,0}^*} \notag \\
     &=1-{2({\rm Im}\,x_{1,0})^2\over  (z-{\rm Re}\,x_{1,0})^2+({\rm Im}\, x_{1,0})^2}-2{\rm i}\,{{\rm Im}\,x_{1,0}\,(z-{\rm Re}\,x_{1,0})\over  (z-{\rm Re}\,x_{1,0})^2+({\rm Im}\, x_{1,0})^2}, \tag{5.7}
     \end{align}
     and  taking into account (5.5), we can recast (5.6) into the form
     $${\bf m}=\cos\,\theta\,{\rm Re}\,B(x-vt)\, {\boldsymbol e}_1+\cos\,\theta\,{\rm Im}\,B(x-vt)\, {\boldsymbol e}_2+\sin\,\theta\,{\boldsymbol e}_3. \eqno(5.8)$$
     This recovers the traveling wave solution of the HWM equation discovered in [1, 2]. 
          
        \begin{figure}[t]
\begin{center}
\includegraphics[width=8cm]{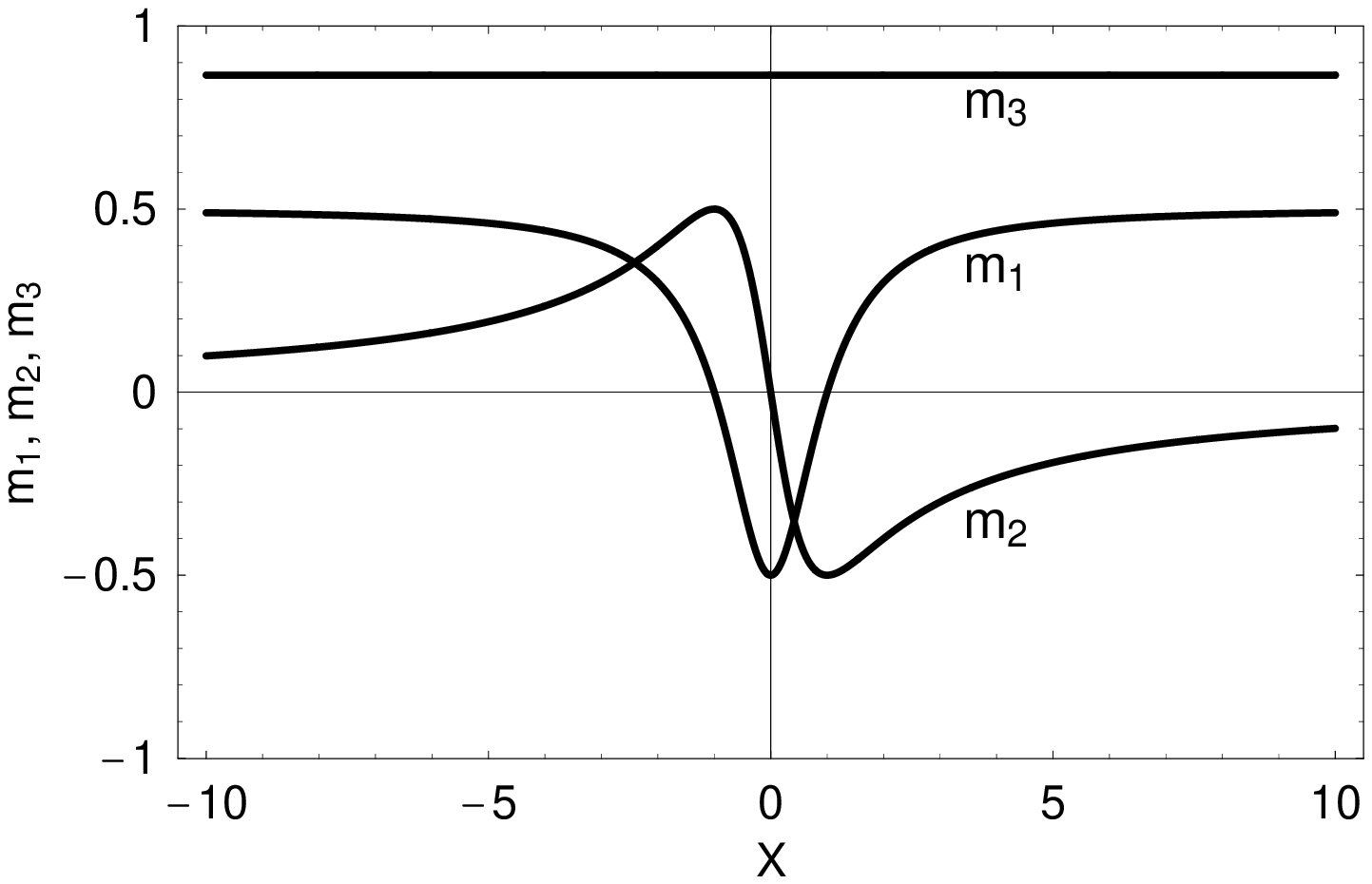}
\end{center}
\noindent {{\bf Fig. 1.}\ A profile of the one-soliton solution as function of the traveling wave coordinate $X=x-vt$  with  the parameters  ${\bf m}_0=(1/2, 0, \sqrt{3}/2), x_{1,0}={\rm i}, {\bf s}_{1,0}=(1/4,{\rm i}/4, 0),
{\bf n}_{1,1}=(1, 0, 0), {\bf n}_{1,2}=(0, 1, 0), \theta=\pi/3\, (v=\sqrt{3}/2)$. }
\end{figure}
     
          A typical profile of the spin  density  ${\bf m}= (m_1, m_2, m_3)$  as  function of the traveling wave coordinate $X=x-vt$  is depicted in Fig. 1. In this example,   it has the form
     $${\bf m}(X)=\left({X^2-1\over 2(X^2+1)}, -{X\over X^2+1}, {\sqrt{3}\over 2}\right). \eqno (5.9)$$
     \par
     An inspection of the component $m_1$ in (5.8) reveals that its amplitude measured from a constant level at spatial infinity  becomes $2\sqrt{1-v^2}/{\rm Im}\,x_{1,0}$ and hence it
     decreases as the velocity increases. 
     This unusual feature can be remedied if one observes the profile of $m_1$
     in the coordinate system moving to the right with a constant velocity $v=1$, for instance.  As a result, the amplitude becomes a monotonically increasing function of the velocity
     in agreement with the velocity-amplitude relation of the soliton.
             \par
              \bigskip
     \noindent {\it 5.2. Two-soliton solution}\par
     \bigskip
   It follows from (4.5) with $N=2$ that the two-soliton solution takes   the form
   \begin{align}
     {\bf m}&={\bf m}_0+{\rm i}\sum_{j=1}^2\left({{\bf s}_j(t)\over x-x_j(t)}-c.c\right) \notag \\
            &={\bf m}_0+{\rm i}\left[{1\over f_2}\{{\bf J}_0\sigma_0 x-{\bf J}_0\sigma_1+{\bf J}_1\sigma_0\}-c.c. \right], \tag{5.10a}
   \end{align}
   where
   $$\sigma_0=1, \quad \sigma_1=(\dot{x}_{1,0}+\dot{x}_{2,0})t+x_{1,0}+x_{2,0}, \eqno(5.10b)$$
   $${\bf J}_0={\bf s}_{1,0}+{\bf s}_{2,0}, 
   \quad {\bf J}_1=(\dot{x}_{1,0}{\bf s}_{1,0}+\dot{x}_{2,0}{\bf s}_{2,0}+x_{1,0}\dot{\bf s}_{1,0}+x_{2,0}\dot{\bf s}_{2,0})t+{x}_{1,0}{\bf s}_{1,0}+{x}_{2,0}{\bf s}_{2,0}, \eqno(5.10c)$$
     $$\dot{x}_{1,0}=-{\rm i}{\bf n}_1\cdot \left({\rm i}{\bf m}_0-{{\bf s}_{2,0}\over x_{1,0}-x_{2,0}}
 +{{\bf s}^*_{1,0}\over x_{1,0}-x^*_{1,0}}+ {{\bf s}^*_{2,0}\over x_{1,0}-x^*_{2,0}}\right),  $$
       $$\dot{x}_{2,0}=-{\rm i}{\bf n}_2\cdot \left({\rm i}{\bf m}_0-{{\bf s}_{1,0}\over x_{2,0}-x_{1,0}}
 +{{\bf s}^*_{1,0}\over x_{2,0}-x^*_{1,0}}+ {{\bf s}^*_{2,0}\over x_{2,0}-x^*_{2,0}}\right),  \eqno(5.10d)$$
 $$\dot{\bf s}_{1,0}=-2\,{{\bf s}_{1,0}\times {\bf s}_{2,0}\over (x_{1,0}-x_{2,0})^2}, \quad   \dot{\bf s}_{2,0}=-2\,{{\bf s}_{2,0}\times {\bf s}_{1,0}\over (x_{2,0}-x_{1,0})^2}. \eqno(5.10e)$$
 Here, ${\bf s}_{j,0}\, (j=1,2)$ are given by (4.15) with (4.26).  \par
 The tau-function $f_2$  from $(2.21b)$ may be written in a convenient form suitable for investigating the asymptotics 
 of the two-soliton solution.  To be more specific,  we put
 \begin{align}
 f_2&= \begin{vmatrix} x-v_1t+\alpha_1 & \beta_{1,2}\\
                       \beta_{2,1}& x-v_2t+\alpha_2 \end{vmatrix} \notag \\
                       &=x^2-\{(v_1+v_2)t-(\alpha_1+\alpha_2)\}x+v_1v_2t^2-(v_2\alpha_1+v_1\alpha_2)t+\alpha_1\alpha_2-\beta_{1,2}\beta_{2,1}, \tag{5.11}
                       \end{align}
 where $v_1, v_2\in {\mathbb R}$ and $ \alpha_1, \alpha_2, \beta_{1,2}, \beta_{2,1}\in {\mathbb C}$ are unknown  parameters.
  To determine these unknowns, we compare 
   $(2.21b)$ with (5.11) and obtain  the following system of algebraic equations:
 $$v_1+v_2=\dot{x}_{1,0}+\dot{x}_{2,0}, \eqno(5.12a)$$
 $$\alpha_1+\alpha_2=-(x_{1,0}+x_{2,0}), \eqno(5.12b)$$
 $$v_1v_2=\dot{x}_{1,0}\dot{x}_{2,0}-{2\,{\bf s}_{1,0}\cdot {\bf s}_{2,0} \over (x_{1,0}-x_{2,0})^2}, \eqno(5.12c)$$
 $$v_2\alpha_1+v_1\alpha_2=-(\dot{x}_{1,0}x_{2,0}+x_{1,0}\dot{x}_{2,0}), \eqno(5.12d)$$
 $$\alpha_1\alpha_2-\beta_{1,2}\beta_{2,1}=x_{1,0}x_{2,0}. \eqno(5.12e)$$
 If we take into account  $(5.12a)$ and $(5.12c)$, we see that   $v_1$ and $v_2$  follow by solving the quadratic  equation for $v$
 $$v^2-(\dot{x}_{1,0}+\dot{x}_{20})v+\dot{x}_{1,0}\dot{x}_{2,0}-{2\,{\bf s}_{1,0}\cdot {\bf s}_{2,0} \over (x_{1,0}-x_{2,0})^2}=0, \eqno(5.13a)$$
 which gives rise to the solutions
 $$v_{1,2}={1\over 2}\left\{\dot{x}_{1,0}+\dot{x}_{2,0}\mp\sqrt{(\dot{x}_{1,0}-\dot{x}_{2,0})^2+{8\,{\bf s}_{1,0}\cdot {\bf s}_{2,0} \over (x_{1,0}-x_{2,0})^2}}\right\}, \eqno(5.13b)$$
 where the plus (minus) sign corresponds to $v_2 (v_1)$. The reality of the velocities $v_1$ and $v_2$ will be shown in remark 3.
 On the other hand, $\alpha_1$ and $\alpha_2$ are found from $(5.12b)$ and $(5.12d)$ to be
 $$\alpha_1={1\over v_1-v_2}\left\{-(x_{1,0}+x_{2,0})v_1+\dot{x}_{1,0}x_{2,0}+x_{1,0}\dot{x}_{2,0}\right\}, \eqno(5.14a)$$
  $$\alpha_2={1\over v_2-v_1}\left\{-(x_{1,0}+x_{2,0})v_2+\dot{x}_{1,0}x_{2,0}+x_{1,0}\dot{x}_{2,0}\right\}. \eqno(5.14b)$$
  Introducing $(5.14)$ with (5.13) into $(5.12e)$, we obtain, after a few computations, the following simple relation
  $$\beta_{1,2}\beta_{2,1}={2\,{\bf s}_{1,0}\cdot {\bf s}_{2,0}\over (v_1-v_2)^2}. \eqno(5.15)$$
  Consequently, we can put
  $$\beta_{1,2}=\beta_{2,1}={\sqrt{2\,{\bf s}_{1,0}\cdot {\bf s}_{2,0}}\over |v_1-v_2|}. \eqno(5.16)$$
  \par
   The new parameterization of the tau-function $f_2$ allows us to  explore the properties of the two-soliton solution.
    A typical example of the two-soliton solution is depicted in Fig. 2 in which  the form of $\bf m$ at $t=0$ is given by
  $${\bf m}(x, 0)=\left(-{4x(x^2-1)\over (x^2+1)(x^2+4)}, {8x^2\over (x^2+1)(x^2+4)}, {x^2-4\over x^2+4}\right).\eqno(5.17)  $$
  The figure displays the head-on collision of two solitons. The solitonic nature of the solution is seen obviously, i.e., 
  each soliton appears without changing its profile after the collision.
    \par
             \begin{figure}[t]
\begin{center}
\includegraphics[width=16cm]{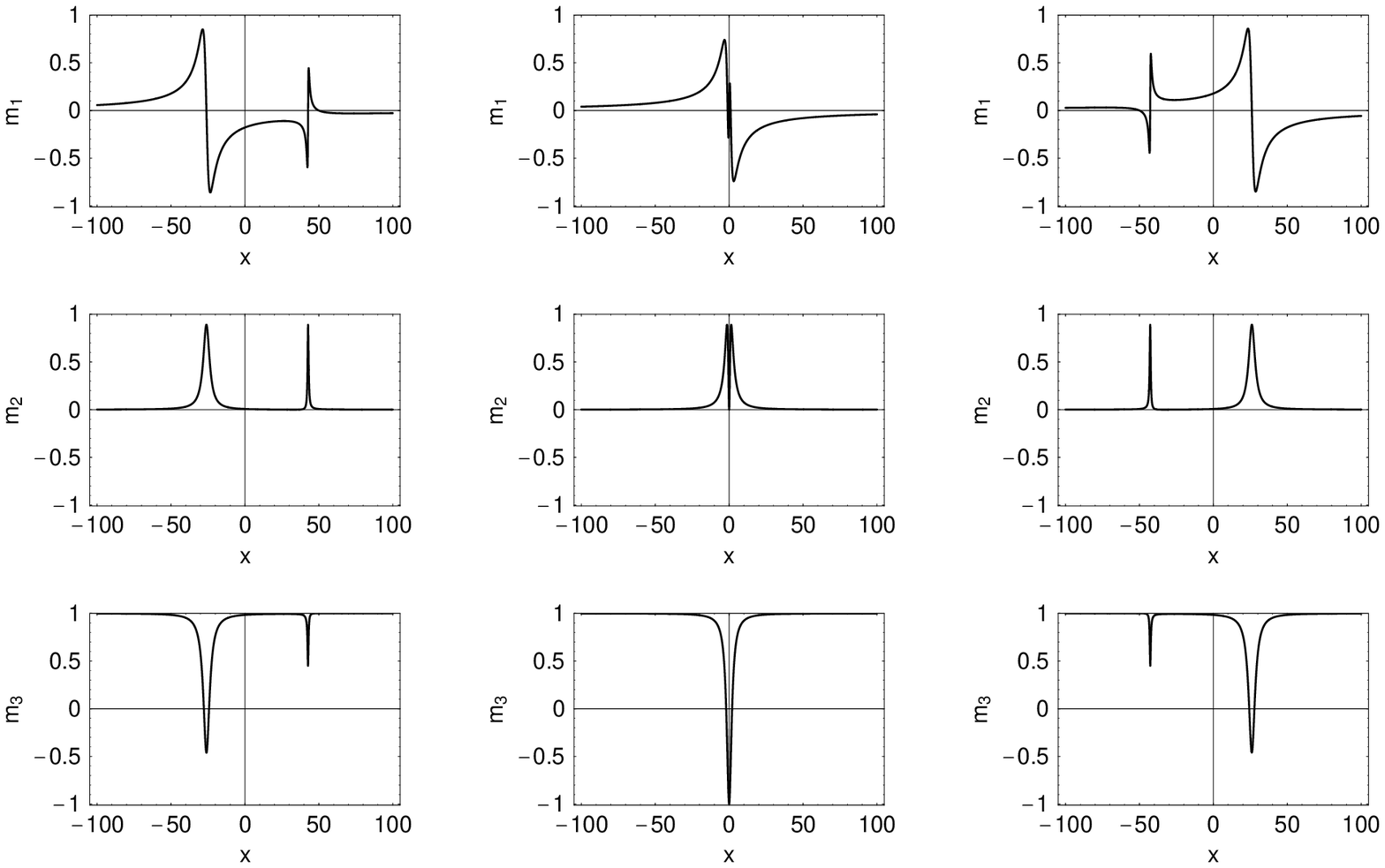}
\end{center}
\noindent {\bf Fig. 2.}\  Profiles of the two-soliton solution at three different times: left panel $t=-50$; middle panel $t=0$; right panel $t=50$.
The parameters are ${\bf m}_0=(0, 0, 1), x_{1,0}={\rm i}, x_{2, 0}=2{\rm i}, {\bf s}_{1,0}=(-4{\rm i}/3, 4/3, 0), {\bf s}_{2,0}=(10{\rm i}/3, -8/3, 2), {\bf n}_{1,1}=(0, -1, 0),
{\bf n}_{1,2}=(1, 0, 0), {\bf n}_{2,1}=(0, 5/4, -3/5), {\bf n}_{2,2}=(-1, 0, 0), v_1=-0.854,  v_2=0.521$. 
\end{figure}
      The detailed information  of the interaction process of  solitons  may be extracted from the 
   motion of poles  in the complex plane.
   This purpose can be attained  in principle by solving the equations of motion (1.4) for $x_1$ and $x_2$.
   When only  solitons are present, however, an alternative way is to find the pole positions  by solving the quadratic equation $f_2=0$.
   Actually, using (5.11) with (5.16),  we find that
   $$x_{1,2}={1\over 2}\left\{(v_1+v_2)t-\alpha_1-\alpha_2\pm \sqrt{D}\right\}, \eqno(5.18a)$$
  where
  $$D=\{(v_1-v_2)t-(\alpha_1-\alpha_2)\}^2+{8\,{\bf s}_{1,0}\cdot {\bf s}_{2,0}\over( v_1-v_2)^2}. \eqno(5.18b)$$
  A simple analysis reveals that the asymptotic forms  of $x_1$ and $x_2$ from (5.18) as $t\rightarrow \pm\infty$ take the same forms and given respectively by
  $$x_1 \sim v_1t-\alpha_1+{2\,{\bf s}_{1,0}\cdot {\bf s}_{2,0}\over (v_1-v_2)^2}\,{1\over (v_1-v_2)t-(\alpha_1-\alpha_2)}, \eqno(5.19a)$$
  $$x_2 \sim v_2t-\alpha_2-{2\,{\bf s}_{1,0}\cdot {\bf s}_{2,0}\over (v_1-v_2)^2}\,{1\over (v_1-v_2)t-(\alpha_1-\alpha_2)}. \eqno(5.19b)$$
   On the other-hand, it follows from (4.30) with $N=2$ that
  $${\bf s}_1={1\over x_2-x_1}\,({\bf J}_0x_2-{\bf J}_1), \eqno(5.20a)$$
  $$ {\bf s}_2={1\over x_1-x_2}\,({\bf J}_0x_1-{\bf J}_1). \eqno(5.20b)$$
  Taking into account $(5.10c)$ and (5.19), one can see that both ${\bf s}_1$ and ${\bf s}_2$ approach the constant values as $t \rightarrow \pm \infty$.
    \begin{figure}[t]
\begin{center}
\includegraphics[width=8cm]{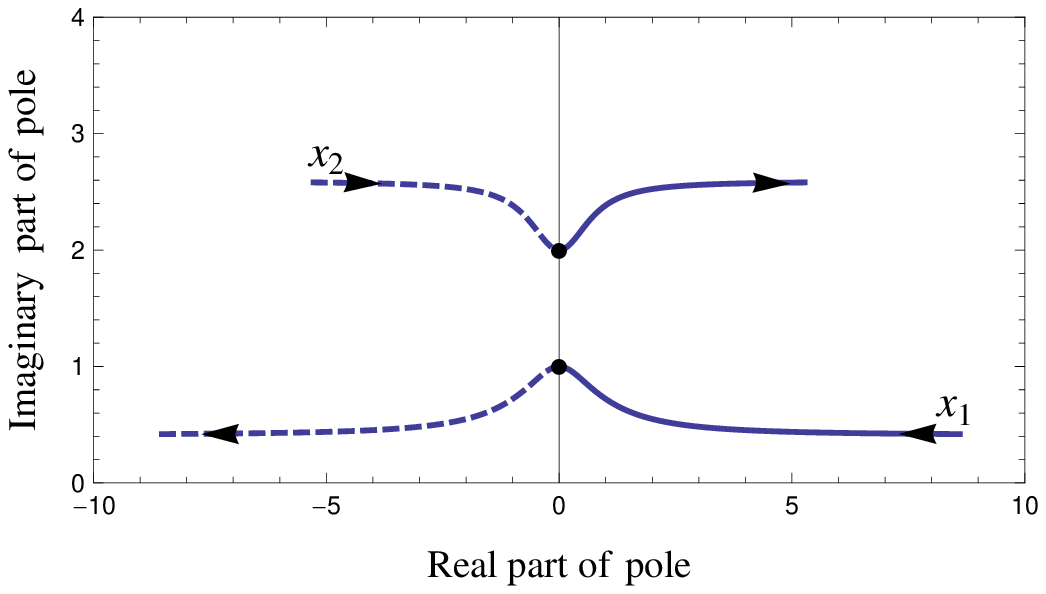}
\end{center}
\noindent {\bf Fig. 3.}\ The time evolution of the poles $x_1$\,(solid line) and $x_2$\,(dotted line) in the time interval $-10\leq t\leq 10$, where the parameters are the same as those  used in Fig. 2. The arrows indicate the 
 direction of motion, and the black dots mark the positions of the poles at $t=0$. In this example, ${\rm Im}\,x_1(\pm\infty)=0.409, {\rm Im}\,x_2(\pm\infty)=2.59$.
  \end{figure}
      \begin{figure}[t]
\begin{center}
\includegraphics[width=8cm]{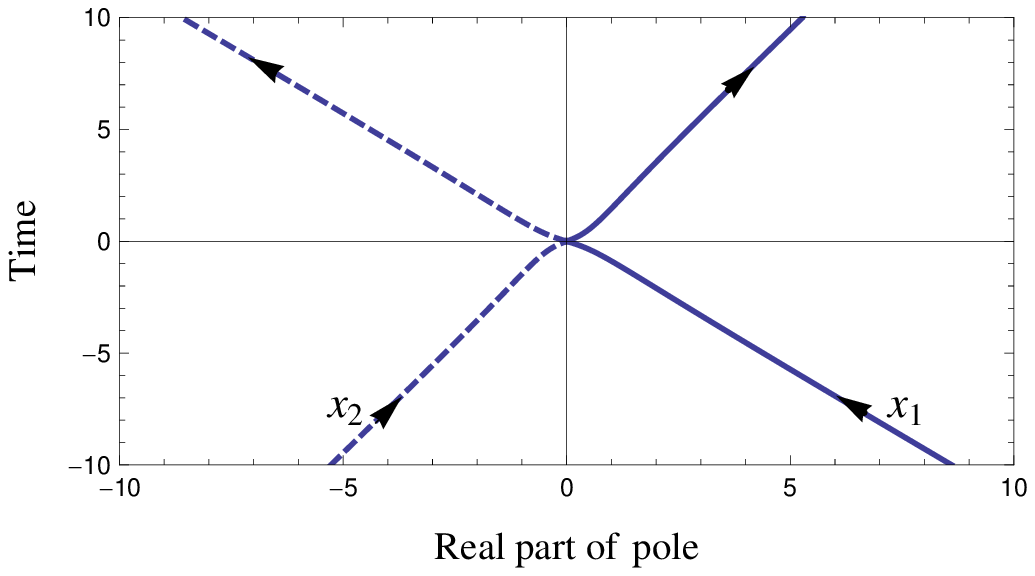}
\end{center}
\noindent {\bf Fig. 4.}\ The time evolution of the real part of poles ${\rm Re}\,x_1$\,(solid line) and ${\rm Re}\,x_2$\,(dotted line) in the time interval $-10\leq t\leq 10$ corresponding to Fig. 3.
  \end{figure}
    Actually, the leading order asymptotics of ${\bf s}_1$ and  ${\bf s}_2$ are found to be
  $${\bf s}_1 \sim {1\over v_2-v_1}\{({\bf s}_{1,0}+{\bf s}_{2,0})v_2-(\dot{x}_{1,0}{\bf s}_{1,0}+\dot{x}_{2,0}{\bf s}_{2,0}+x_{1,0}\dot{\bf s}_{1,0}+x_{2,0}\dot{\bf s}_{2,0})\}, \eqno(5.21a)$$
  $${\bf s}_2 \sim {1\over v_1-v_2}\{({\bf s}_{1,0}+{\bf s}_{2,0})v_1-(\dot{x}_{1,0}{\bf s}_{1,0}+\dot{x}_{2,0}{\bf s}_{2,0}+x_{1,0}\dot{\bf s}_{1,0}+x_{2,0}\dot{\bf s}_{2,0})\}. \eqno(5.21b)$$
  \par
  The asymptotic form of ${\bf m}$ from (5.10)  is  obtained by introducing (5.19) and (5.21) into it, giving
  $${\bf m} \sim {\bf m}_0+{\rm i}\sum_{j=1}^2\left({{\bf s}_j(\infty)\over x-v_jt+\alpha_j}-c.c. \right), \eqno(5.22)$$
  where  ${\bf s}_j(\infty)\, (j=1, 2)$ stand for the  asymptotic values given by (5.21). The above expression shows clearly that the two-soliton solution  is composed  of a superposition  of  single solitons 
  given by (5.4).  Moreover,  it exhibits no phase shifts after the collision.
  This remarkable feature is common to that of the rational (or algebraic) multisoliton solutions of the BO and nonlocal NLS equations [18-21]. 
  However, to the best of our knowledge, it is the first example observed in the {\it head-on}  collision of rational  solitons.
  Recall in this respect that the BO rational solitons exhibit no phase shifts after {\it overtaking} collisions [18].
  Fig. 3 shows the time evolution of the poles $x_1$ and $x_2$ in the complex plane corresponding to the two-soliton solution depicted in Fig. 2.
   One can observe that the distance between two poles becomes minimum at $t=0$ at which  instant two solitons would collide.
    See  the corresponding profiles of solitons in Fig. 2. 
  A detailed inspection of the expressions of the poles  $x_1$ and $x_2$ from (5.18) indicates that the imaginary part of each pole exhibits a discontinuity at $t=0$
  in such a way that the poles interchange their imaginary parts. 
  This intriguing feature is sometimes called an exchange of identity.  Fig. 4 depicts the time evolution of the real part of poles. While their trajectories  are continuous  during their motion, an exchange of
  identity occurs at the instant of the collision of poles. 
  \par
           \bigskip
     \noindent {\it 5.3. N-soliton solution}\par
     \bigskip
 The construction of the $N$-soliton solution can be done following  a purely algebraic procedure developed in Section 4. 
 Here, we discuss the property of the $N$-soliton solution focusing on its asymptotic behavior. 
 Now, we assume the tau-function $f_N$ of the determinantal form
 $$f_N=\Bigl|\Bigl(\delta_{jk}(x-v_jt+\alpha_j)+(1-\delta_{jk})\beta_{j,k}\Bigr)_{1\leq j, k\leq N}\Bigr|, \quad (\beta_{j,k}=\beta_{k,j}), \eqno(5.23)$$
 which is a natural generalization of the tau-function of the two-soliton solution (5.11).  We compare  (2.20) with (5.23)
 and see that equating the coefficients of the terms $t^sx^j\, (j=0, 1, ..., N-1; s=0, 1, ..., N)$  
 yields the $N(N+3)/2$ equations for the unknowns $v_j, \alpha_j\, (j=1, 2, ..., N)$ and $\beta_{j,k}\, (j,k=1,2, .., N; j\not=k)$.
 Since the number of the unknowns is equal to that of the equations, we can determine these unknowns in principle.  
 An explicit computation has been performed for the two-soliton solution, and  as shown in (5.12), the five equations were derived for the same number of unknowns.
 \par
 Let us now investigate the  behavior of the solution.  An asymptotic analysis using (5.23) reveals that the poles $x_j$ behave like
 $$x_j \sim v_jt-\alpha_j, \quad t\rightarrow \pm\infty, \quad (j=1, 2, ..., N).\eqno(5.24)$$
 Referring to (5.24), we find from  (4.30) with (4.6) that  the spin variables ${\bf s}_j$ approach the constant vectors as $t\rightarrow \pm\infty$. 
 The asymptotic form of ${\bf m}$ then becomes
   $${\bf m} \sim {\bf m}_0+{\rm i}\sum_{j=1}^N\left({{\bf s}_j(\infty)\over x-v_jt+\alpha_j}-c.c. \right). \eqno(5.25)$$
This represents simply a superposition of $N$ single solitons, each of which undergoes no phase shifts after collisions.  \par
\bigskip
\noindent {\bf Remark 3.} The parameter $v_j$ in the asymptotic expression (5.24)  is the velocity of the $j$th pole and hence it should be a
real quantity.  In the case of the two-soliton solution, it is given by $(5.13b)$.  Here, we provide a
general proof that $v_j$ is real and satisfies the inequality $|v_j|<1$. In addition, we show that ${\rm Im}\, \alpha_j\not=0$. 
We first recall that Eq. (1.5) is satisfied for $t>0$ [10].  Taking the limit $t\rightarrow \infty$  for  the expression of $\dot x_j(t)$ and then
substituting the asymptotic form of $x_j$ from (5.24) under the assumptions $v_j\not=v_k$ for $j\not=k$ and $v_j, \alpha_j\in \mathbb{C}$, we obtain
$$v_j={\rm i}\,{{\bf m}_0\cdot({\bf s}_j(\infty)\times {\bf s}_j^*(\infty)) \over {\bf s}_j(\infty)\cdot {\bf s}_j^*(\infty)}, \quad (j=1, 2, ..., N). \eqno(5.26)$$
We see from this expression that $v_j^*=v_j$, implying that $v_j$ is real.  If we use the formula $(2.6b)$  subjected to  the conditions ${\bf m}_0^2=1, {\bf s}_j^2(\infty)=0$, 
we can derive the relation
$$\{{\bf m}_0\cdot({\bf s}_j(\infty)\times {\bf s}_j^*(\infty)\}^2=2({\bf m}_0\cdot{\bf s}_j(\infty))({\bf m}_0\cdot{\bf s}_j^*(\infty))({\bf s}_j(\infty)\cdot{\bf s}_j^*(\infty))
-({\bf s}_j(\infty)\cdot{\bf s}_j^*(\infty))^2. \eqno(5.27)$$
It follows from (5.26) and (5.27) that
$$v_j^2=1-{2({\bf m}_0\cdot{\bf s}_j(\infty))({\bf m}_0\cdot{\bf s}_j^*(\infty))\over{\bf s}_j(\infty)\cdot{\bf s}_j^*(\infty)}, \quad (j=1, 2, ..., N). \eqno(5.28)$$ 
Let $v_j^2=1-\delta_j$ where $\delta_j$ stands for the second term on the right-hand side of (5.28).  The inequality $0<\delta_j\leq 1$  follows from (5.27) if ${\bf m}_0^2=1, {\bf s}_j^2(\infty)=0$.
Plugging this inequality into (5.28),  we finally  arrive at the  result $|v_j|<1\ (j=1, 2, ..., N)$.  A similar argument using the large time analog of (1.6) yields the relation
$${\rm Im}\, \alpha_j=-{1\over 2}\,{{\bf s}_j(\infty)\cdot{\bf s}_j^*(\infty)\over {\bf m}_0\cdot{\bf s}_j(\infty)}, \quad (j=1, 2, ..., N). \eqno(5.29)$$
It turns out that ${\rm Im}\, \alpha_j\not=0$ and its sign is determined by the sign of the term ${\bf m}_0\cdot{\bf s}_j(\infty)$. 
The basic assumption in deriving the system of equations  with constraints (1.3)-(1.6) is  that ${\rm Im}\,x_j(t)>0\ (j=1, 2, ..., N)$ for $ t>0$.  
This implies that the tau-function  $f_N(x,t)$ from (2.19)  never becomes zero for real $x$.  The proof of this fact has not been established as yet
even though it can be checked  numerically for $N=2, 3$. However, the asymptotic form (5.24) of $x_j$  with ${\rm Im}\,\alpha_j\not=0$  strongly
supports the above statement.  This important problem should be addressed in a future study.
\par
\bigskip
\noindent {\bf 6. Concluding remarks}\par
\bigskip
In this paper, we  found  an explicit Lax pair  for a many-body dynamical system associated with the HWM equation.
Although the dynamical system has been  found to be equivalent to the spin CM system [10], its Lax pair was presented here for the first time, in which
a key identity (2.5) played a central role.
We also derived  the conservation laws of the system using the Lax pair following a standard procedure, and clarified the underlying
Hamiltonian structure.   We stress that  analytical expressions of the multisoliton solutions of the HWM equation were also presented  for the first time,
even though a numerical scheme for obtaining them has been developed in [10].  A new parameterization of the $N$ soliton 
tau-function enables us to explore its large time asymptotics,  showing that no phase shifts appear after the head-on collisions of solitons. 
 The study of the HWM equation just begins and  a number of problems remain unsolved.
In conclusion, we discuss some of them.  \par
\medskip
\noindent 1. One of the most important issues  will  be the analysis of the HWM equation by means of the inverse scattering transform method (IST).
Although the IST has been used  successfully to  the local nonlinear evolution equations such as the KdV and NLS equations,  its application to
 nonlocal nonlinear equations  is far from satisfactory.  See  [22-25]  for the BO and nonlocal NLS equations.
The similar situation happens to the HWM equation for which the Lax pair takes a nonlocal form [7]. 
Nevertheless, the nonlocal Riemann-Hilbert approach may work effectively for these nonlocal equations. See, for instance,  [26].
\par
\noindent 2.  The HWM equation is formally  shown to exhibit an infinite number of conservation laws including 
the mass, momentum and energy [1, 2].  However, the explicit forms of the higher conservation laws in terms of the spin densities have not been derived  yet.
In this respect, it should be noted that the information of the conservation laws of the spin CM system derived here has no direct relevance to that  of the
HWM equation. \par
\noindent 3. The HWM equation is obtained formally from the classical Heisenberg ferromagnet equation ${\bf m}_t={\bf m}\times {\bf m}_{xx}$  if one replaces
an $x$ derivative  by the Hilbert transform. This formal derivation is just similar to the relation between the KdV and BO equations.
A number of outcomes have been obtained for the Heisenberg ferromagnet equation [27].  Their analogs for the HWM equation will be worth studying. 
For instance, the Heisenberg ferromagnet equation is known to be gauge-equivalent to the NLS equation [28-30] and hence the HWM equation may be expected to have
 a gauge-equivalent nonlocal nonlinear equation. \par
\noindent 4. There exist several exact methods of solutions for solving the soliton equations. Among them, the direct method  provides a very powerful mean to construct
multisoliton solutions [31, 32]. In fact, it has been applied to the BO and nonlocal NLS equations to obtain the explicit $N$-soliton formulas [18-21, 33].
On the other hand,  the pole expansion method is rather sophisticated because it needs to solve the equations of motion for the corresponding dynamical system, as exemplified in this paper.
This situation will be apparent if one compares the derivation of the $N$-soliton solution of the nonlocal NLS equation, for instance  by means of the pole expansion method [16] and direct method [21].  
An application of the direct method to the HWM equation is a challenging issue. 
\par
\noindent 5.  While we were concerned with soliton solutions on the real line, the construction of periodic solutions is an important issue.  Both the pole expansion method and direct method 
are amenable to this problem.  An application of the former method to the nonlocal NLS equation has already been developed, whereby  the quantities  analogous to (4.4) were used effectively [16].
We will report the results associated with periodic solutions of the HWM equation in a subsequent paper. \par

\par
\bigskip
\noindent{\bf Appendix A. Proof of Proposition 4} \par
\bigskip
The proof of (4.7) can be performed by comparing the coefficients of $\epsilon^n$ on both sides. Explicitly, it yields
$$n\dot{X}X^{n-1}=BX^n-X^nB+\sum_{l=0}^{n-1}X^lLX^{n-l-1}. \eqno (A.1)$$
The proof of $(A.1)$ proceeds by the mathematical induction.  For $n=1$, $(A.1)$ becomes 
$$\dot{X}=BX-XB+L, \eqno(A.2)$$
which is just (2.14).  Assume that $(A.1)$ holds for $n=m(\geq 2)$, which, multiplied  by $X$ from the right, gives
$$m\dot{X}X^{m}=BX^{m+1}-X^mBX+\sum_{l=0}^{m-1}X^lLX^{m-l}. \eqno(A.3)$$
If we introduce  the term $BX$ from $(A.2)$ into the second term on the right-hand side of $(A.3)$, and using an obvious relation $X\dot{X}=\dot{X}X$
which follows since $X$ is a diagonal matrix, we obtain
$$(m+1)\dot{X}X^{m}=BX^{m+1}-X^{m+1}B+\sum_{l=0}^{m}X^lLX^{m-l}, \eqno(A.4)$$
showing that $(A.1)$ holds for $n=m+1$.\quad 
$\Box$
\par
\bigskip
\noindent{\bf Appendix B. Proof of Proposition 5} \par
\bigskip
It follows by differentiating $K=YL$  by $t$ and using (2.1) and (4.7) that
\begin{align}
\dot{K}&=(BY-YB+\epsilon YLY)L+Y(BL-LB) \notag \\
&=BYL+\epsilon YLYL-YLB \notag \\
&=BK-KB+\epsilon K^2. \notag
\end{align}
$\Box$
\par
\bigskip
\newpage
\noindent{\bf Appendix C. Proof of Proposition 6} \par
\bigskip
 We  differentiate $\mathscr{P}_n={\rm Tr}(S^2K^nY)$ by $t$ and then use the time derivative $\dot{S}$ from  (2.11), $\dot{Y}$ from  $(4.7)$ with $K=YL$ and $\dot{K}$ from
$(4.8)$, respectively.  This  leads to 
\begin{align}
\dot{\mathscr{P}}_n&= {\rm Tr}\Bigl[(BS^2-S^2B)K^nY+S^2\sum_{l=0}^{n-1}K^l(BK-KB+\epsilon K^2)K^{n-l-1}Y \notag \\
&+S^2K^n(BY-YB+\epsilon YLY)\Bigr]\notag \\
&={\rm Tr}\left[S^2\left\{\left(\sum_{l=0}^nK^lBK^{n-l}-\sum_{l=-1}^{n-1}K^{l+1}BK^{n-l-1}\right)Y+\epsilon (n+1) K^{n+1}Y\right\}\right] \notag \\
&=\epsilon (n+1) {\rm Tr}(S^2K^{n+1}Y) \notag \\
&=\epsilon (n+1)\mathscr{P}_{n+1}. \notag
\end{align}
$\Box$
\par
\bigskip
\noindent{\bf Appendix D. Proof of Proposition 7} \par
\bigskip
The proof proceeds by the mathematical induction.  For $n=1$, $(4.10)$ reduces to
$$\sum_{l=1}^\infty \epsilon^l\,{ d\mathscr{Q}_l\over dt}=\epsilon {\rm Tr}(S^2KY), \eqno(D.1)$$
which  is verified as follows:  First, referring to the definition $\mathscr{Q}_l={\rm Tr}(S^2X^l)$, one has
$$\sum_{l=1}^\infty \epsilon^l\mathscr{Q}_l={\rm Tr}\left\{S^2(Y-I)\right\}. \eqno(D.2)$$
Then, differentiation this expression by $t$ yields
$$\sum_{l=1}^\infty \epsilon^l{d \mathscr{Q}_l\over dt}={\rm Tr}\left\{(\dot{S}S+S\dot{S})(Y-I)+S^2\dot{Y}\right\}. \eqno(D.3)$$
Last, substituting $\dot{S}$ from (2.11) and $\dot{Y}$ from $(4.7)$ into $(D.2)$, we deduce
\begin{align}
\sum_{l=1}^\infty \epsilon^l{d \mathscr{Q}_l\over dt} &={\rm Tr}\left[S^2\left\{(Y-I)B-B(Y-I)+BY-YB+\epsilon YLY\right\}\right]  \notag \\
&=\epsilon {\rm Tr}(S^2KY). \tag{D.4}
\end{align}
   which is  $(D.1)$. 
   Assume that $(4.10)$ holds for $n=m(>1)$, i.e.,
   $$\sum_{l=m}^\infty \epsilon^l\,{d^m \mathscr{Q}_l\over dt^m}=\epsilon^mm!{\mathscr P}_m. \eqno(D.5)$$
   We differentiate  $(D.5)$ by $t$ and use $(4.9)$ to obtain
   $$\sum_{l=m}^\infty \epsilon^l\,{d^{m+1} \mathscr{Q}_l\over dt^{m+1}}=\epsilon^{m+1}(m+1)!{\mathscr P}_{m+1}. \eqno(D.6)$$
   Comparison of the coefficients of $\epsilon^{m}$  on both sides of $(D.6)$ yields the relation $d^{m+1}\mathscr{Q}_m/dt^{m+1}=0$.
   It turns out that  $(D.6)$ reduces to   
   $$\sum_{l=m+1}^\infty \epsilon^l\,{d^{m+1} \mathscr{Q}_l\over dt^{m+1}}=\epsilon^{m+1}(m+1)!{\mathscr P}_{m+1}, \eqno(D.7)$$
   implying that $(4.10)$ holds for $n=m+1$. \quad $\Box$. \par
   \bigskip
   \noindent{\bf Acknowledgements} \par
   \bigskip
   The author is grateful to two anonymous reviewers for their useful comments and suggestions. 
   After the acceptance of the paper for publication, Professor Langmann informed me that an integrable generalization of the HWM equation
   was proposed which is related to the $A$-type hyperbolic spin Calogero-Moser system. See B.K. Berntson, R. Klabbers and E. Langmann, The non-chiral intermediate Heisenberg
   ferromagnet equation, arXiv preprint: 2110.06239 (2021). 
   \par
   \newpage
  \leftline{\bf References} \par
\baselineskip=5.5mm
\begin{enumerate}[{[1]}]
\item T. Zhou and M. Stone, Solitons in a continuous classical Haldane-Shastry spin chain, Phys. Lett. A 379 (2015) 2817-2825.
\item  E. Lenzmann and A. Schikorra, On energy-critical half-wave maps into ${\mathbb S}^2$, Inv. Math. 213 (2018) 1-82.
\item E. Lenzmann and J. Sok, Derivation of the half-waves maps equation from Calogero-Moser spin system, arXiv. 2007.15323 (2020).
\item J. Gibbons and T. Hermsen, A generalization  of the Calogero-Moser system, Physica D11 (1984) 337-348.
\item  S. Wojciechowski, An integrable marriage of the Euler equations with the Calogero-Moser system, Phy. Lett. 111A (1985) 101-103.
\item  I. Krichever, O. Babelon, E. Billey and M. Talon, Spin generalization of the Calogero-Moser system and the matrix KP equation,
in: Topics in Topology and Mathematical Physics, vol. 170 ed. S. Novikov, American Mathematical Society,  Providence RI, 1995, 83-120.
\item  P. G\'erard and E. Lenzmann, A Lax pair structure for the half-wave maps equation, Lett. Math. Phys. 108 (2018) 1635-1648.
\item E. Lenzmann, A short primer on the half-wave maps equation, Journ\'ees \'Equations aux d\'eriv\'ees partielles (2018) 1-12.
\item M. D. Kruskal, The Korteweg-de Vries equation and related evolution equations, Lect. Appl. Math. 15 (1974) 61-83.
\item B. K. Berntson, R. Klabbers and E. Langmann,  Multi-solitons of the half-wave maps equation and Calogero-Moser spin-pole dynamics, J. Phys. A: Math. Theor. 52 (2020) 505702 (32pp).
\item H. Airault, H. P. Mckean and J. Moser, Rational and elliptic solutions of the Korteweg- de Vries equation and a related many-body problem, Comm. Pure and Appl. Math. 30 (1977) 95-148.
\item D. V. Choodnovsky and G. V. Choodnovsky, Pole expansion of nonlinear partial differential equations,  Il Nuovo Cimento  B 40 (1977) 339-353
\item K. M. Case, The $N$-soliton  solution of the Benjamin-Ono equation, Proc. Natl. Acad. Sci. 75 (1978) 3562-3563. 
\item H. H. Chen, Y. C. Lee and N. R. Pereira, Algebraic internal wave solitons and integrable Calogero-Moser-Sutherland $N$-body problem, Phys. Fluids 22 (1979) 187-188. 
\item M. A. Olshanetsky and A. M. Perelomov, Classical integrable finite-dimensional systems related to Lie algebras, Phys. Rep. 71 (1981) 313-400.
\item  Y. Matsuno,  Calogero-Moser-Sutherland dynamical systems associated with nonlocal nonlinear Schr\"odinger equation for envelope waves, J. Phys. Soc. Jpn. 71 (2002) 1415-1418.
\item  S. Wojciechowski, Superintegrability of the Calogero-Moser system, Phy. Lett. 95A (1983)  279-281.
\item  Y. Matsuno,  Exact multi-soliton solution of the  Benjamin-Ono equation,  J. Phys. A: Math. Gen. 12 (1979) 619-621.
\item Y. Matsuno, Interaction of the Benjamin-Ono solitons, J. Phys. A: Math. Gen. 13 (1980) 1519-1536.
\item Y. Matsuno,  Dynamics of interacting algebraic solitons, Int. J. Mod. Phys. B9 (1995) 1985-2081.
\item Y. Matsuno, Multiperiodic and multisoliton solutions of a nonlocal nonlinear Schr\"odinger equation for envelope waves,  Phys. Lett. A 278 (2000)53-58.
\item  Y. Matsuno, Exactly solvable eigenvalue problems for a nonlocal nonlinear Schr\"odinger equation for envelope waves,  Inverse Problems 18 (2002) 1101-1125.
\item Y. Matsuno, A Cauchy problem for the nonlocal nonlinear Schr\"odinger equation, Inverse Problems 20 (2004) 437-445.
\item Y. Matsuno,  Recent topics on a class of nonlinear integrodifferential evolution equations of physical significance, in: Advances in Mathematics Research,
  vol.4 ed. G. Oyibo, Nova Science, New York,  2003, 19-89.
\item  J.-C Saut, Benjamin-Ono and intermediate long wave equations: modeling, IST and PDE, in: 
       Nonlinear Partial Differential Equations and Inverse Scattering, 
       Fields Institute Communications vol. 83 eds. P. Miller, P. Perry, J.-C Saut and C. Sulem, Springer, New York,  NY. 2019, 95-160.
\item P. M. Santini,Integrable singular integral evolution equations, in:Important Developments 
             in Soliton Theory, Springer series in Nonlinear Dynamics, eds. A. S. Fokas and V. E. Zakharov, Springer, New York, NY. 1993, 147-177.
\item M. Lakshmanan, The fascinating world of the Landau-Lifshitz-Gilbert equation: an overwiew, Phil. Trans. R. Soc. A 369 (2011) 1280-1300.
\item M. Lakshmanan, Continuous spin system as an exactly solvable dynamical system, Phys. Lett. 61A (1977) 53-54.
\item L. A. Takhtajan, Integration of the continuous Heisenberg spin chain through the inverse scattering method, Phys. Lett. 64A (1977) 235-237.
\item V. E. Zakharov and L. A. Takhtadzhyan, Equivalence of the nonlinear Schr\"odinger equation and the equation of a Heisenberg ferromagnet,  Theor. Math. Phys.  38 (1979) 17-23.
\item Y. Matsuno, Bilinear Transformation Method, Academic, New York, 1984.
\item R. Hirota, The Direct Method in Soliton Theory, Cambridge University Press, Cambridge 2004.
\item Y. Matsuno,  A direct proof of the $N$-soliton solution of the Benjamin-Ono equation by means of Jacobi's formula, J. Phys. Soc. Jpn. 57 (1988) 1924-1929.

\end{enumerate} 
\end{document}